\documentclass[12pt]{article}

\def\be{\begin{equation}}
\def\ee{\end{equation}}
\def\bea{\begin{eqnarray}}
\def\eea{\end{eqnarray}}

\usepackage{amssymb,latexsym}
\usepackage{graphicx}
\usepackage{amssymb, amsmath, amsopn, amsthm}
\usepackage[section]{placeins}

\makeatletter \@addtoreset{equation}{section}
\makeatother

\begin{document}
\begin{titlepage}
\thispagestyle{empty}
\hskip 1 cm
\vskip 0.5cm

\vspace{25pt}
\begin{center}
    { \LARGE{\bf Sinks in the Landscape, Boltzmann Brains, }}\\ \vskip 3mm

 
  { \LARGE{\bf and the Cosmological Constant Problem}\footnote{\it To the memory of Eugene Feinberg, who was trying to make a bridge between science, philosophy and art.}}\\ \vskip 3mm

    \vspace{33pt}

  {\large  {\bf  Andrei Linde}}

    \vspace{10pt}

    \vspace{10pt} {Department of Physics,
    Stanford University, Stanford, CA 94305}

    \vspace{20pt}
 \end{center}

\begin{abstract}

This paper extends the recent investigation of the string theory
landscape   \cite{Ceresole:2006iq}, where  it was found that the
decay rate of dS vacua to a collapsing space with a negative vacuum
energy can be quite large. The parts of space that experience a
decay to a collapsing space, or to a Minkowski vacuum, never return
back to dS space.  The channels of irreversible vacuum decay serve
as sinks for the probability flow. The existence of such sinks is a
distinguishing feature of the string theory landscape. We describe
 relations  between several different probability measures for
eternal inflation taking into account the existence of the sinks.
The local (comoving) description of the inflationary multiverse
suffers from the so-called Boltzmann brain (BB) problem unless the
probability of the decay to the sinks  is sufficiently large. We
show that some versions of the global (volume-weighted) description
do not have this problem even if one ignores the existence of the
sinks. We  argue that if the number of different vacua in the landscape is large enough, the anthropic solution of the  cosmological constant problem in the string landscape scenario should be valid for a broad class of  the probability measures which  solve the BB problem. If this is correct, the solution of the cosmological constant problem may be essentially measure-independent. Finally, we describe a simplified approach to the calculations of anthropic probabilities in the landscape, which is less ambitious but also less ambiguous than other methods.

\end{abstract}

\vspace{10pt}
\end{titlepage}

\tableofcontents

\section{Introduction}

For many decades people have tried to explain strange correlations
between the properties of our universe, the masses of elementary
particles, their coupling constants, and the fact of our existence.
We know that we could not live in a 5-dimensional universe, or in a
universe where the electromagnetic coupling constant, or the masses
of  electrons and protons would be just a few times greater or
smaller than their present values. These and other similar
observations have formed the basis for the anthropic principle.
However, for a long time many scientists  believed that the universe
was given to us as a single copy, and therefore speculations
about these magic coincidences could not have any scientific
meaning.

The situation changes dramatically with the invention of
inflationary cosmology. It was realized that inflation may divide
our universe into many exponentially large domains corresponding to
different metastable vacuum states, forming a huge inflationary
multiverse \cite{linde1982,nuff,Eternal,book}. The total number of
such vacuum states in string theory can be enormously large
\cite{Lerche:1986cx,Bousso:2000xa,Douglas}. A combination of these
two facts with the KKLT mechanism of vacuum stabilization
\cite{Kachru:2003aw} recently gave rise to what is now called the
string landscape scenario \cite{Susskind:2003kw}. Some people like
the new emerging picture of the multi-faceted universe, some people
hate it, but it does not seem that we have much choice in this
matter: We must learn how to live with this new scientific paradigm.

The first step in this direction is to find out which vacua are
possible in string theory and describe their typical properties
etc. \cite{Douglas}. The second step is to find whether different
vacua can coexist side by side in the same universe, separated by
domain walls \cite{Ceresole:2006iq}. Then we need to study the
cosmological evolution during eternal inflation, which would provide
us with a possible map of the multiverse \cite{LLM}.

The final step is the most ambitious and difficult: We want to
find our own place in the landscape and explain the properties of
our part of the universe. The original goal formulated in
\cite{LLM,Bellido,Mediocr} was to find the place where most of the
observers live.  But  an eternally inflating universe is infinite,
so if we study the global structure of the universe and compare
volumes, we are faced with the problem of comparing infinities.
Several different ways of regulating these infinities have been
proposed. Unfortunately, the results of all of these procedures
depend on the prescription for the cutoff
\cite{LLM,Bellido,Mediocr,Vilenkin:2006qf,LMprob,Tegmark:2004qd,Garriga:2005av}.

An alternative possibility is to study an individual observer,
ignoring the volume-related effects and the global structure of the
universe. This is sometimes called ``the local description.'' One
can do this using comoving coordinates,  which are not expanding
during eternal inflation
\cite{Starobinsky:1986fx,Goncharov:1987ir,Garriga:1997ef,Bousso:2006ev}.
This description, unlike the previous ones,  does not tell us much
until we solve the problem of initial conditions. Finally, one may
try to use the methods of Euclidean quantum gravity, see e.g.
\cite{Hawking:2006ur}. However,  this approach is insufficiently
developed. The debates about the Hartle-Hawking wave function
\cite{Hartle:1983ai} versus the tunneling wave function
\cite{Linde:1983mx} have continued for more than 20 years. The
related conceptual problems are extremely complicated despite the
fact  that these wave functions were calculated in the simplest
(minisuperspace) approximation. This approximation, by construction,
does not allow us to study the global structure of an eternally
inflating universe.

The purpose of this paper is to clarify some features of the
landscape in the simplest possible way. Our description will be
incomplete, it will not cover some of the interesting recent
proposals, but we hope that it will be useful anyway. To begin with,
we will concentrate on  drawing several sketches of ``the map of the
universe.'' There are many ways to do this. Each one provides us
with a complementary view on the structure of the universe, and each
of them can be useful. The problems begin if we start using our maps
in an attempt to understand why it  is that   we live in this
particular place at this particular time.

The easiest route to avoid these problems would be to
concentrate on the conditional probabilities; see a discussion in
Sect. \ref{disc}. On the other hand, it would be nice to demonstrate
that even though the part of the inflationary multiverse where
we live is not unique, it is the best, or at least the most probable
one. Only if all our attempts to put us to the ``center of the
universe'' fail,  we will have a right to say, following Copernicus,
that we just happen to live in a not very special part of the
multiverse; perhaps not the best or the worst, maybe not even
close to a maximum of the probability distribution, but just in some
place consistent with our existence.

One way to analyze these issues  is to consider the probability
measure as a part of the theory, and to compare its predictions with
observations. If some of the probability measures lead to obviously
incorrect predictions, we will concentrate on the remaining ones,
which will reduce the uncertainty.

For example, recently it was argued that the Hartle-Hawking wave
function predicts that most of the observes should  exist in a form
of short-living brains (Boltzmann brains, or BB) created by quantum
fluctuations and floating in an empty de Sitter vacuum
\cite{Page:2006dt,Page:2006hr}. More generally, we are talking about a possibility that the local conditions required for the existence of our life (planets, of solar systems, or isolated galaxies) were created by incredibly improbable quantum fluctuations in an empty dS space, instead of being produced in a regular way after  the post-inflationary reheating of the universe. This possibility would contradict observational data.

The Boltzmann brane concept was introduced in \cite{Albrecht:2004ke}, where some possible ways to resolve the related problems were proposed. It was closely related to the ideas developed in Ref. \cite{Dyson:2002pf}.
Among the ``best'' ways to resolve the BB problem suggested in
\cite{Page:2006dt,Page:2006hr} was the prediction of a  doomsday in
$10^{10}$ years from now,  which requires the existence of
superheavy gravitinos. If this is the case, a discovery of
supersymmetric particles at LHC would give us a chance to test the
wave function of the universe and to learn something about our
future.

Using closely related arguments, but without assuming the validity
of the Hartle-Hawking wave function, recently it was claimed that
all attempts at a global description of our universe lead to an
invasion of  Boltzmann brains \cite{Bousso:2006xc}. And since none of
us wants to believe that he or she is a BB, then, according to \cite{Bousso:2006xc}, we must conclude
that all attempts of a global description of the universe should be
abandoned in favor of the particular version of the local description, which was proposed in  \cite{Bousso:2006ev} and called holographic. The
relation of the method proposed in \cite{Bousso:2006ev} to the previously developed methods was not immediately obvious. It was criticized in \cite{VilMulti}, where it was concluded that ``for someone not initiated in holography, this view is very hard to adopt.'' So if the only BB-free prescription is bad, does it mean
that all good prescriptions predict Boltzmann brains all the way
down?

In this paper we will try  to discuss  related issues and analyze
some of the  existing problems. In Section  \ref{dS} we will
describe the  theory of tunneling and quantum diffusion between
different de Sitter vacua. However, this theory only partially
describes the mechanism of the population of the landscape.
According to \cite{Kachru:2003aw}, all dS vacua in the string
landscape scenario are unstable with respect to decay to a Minkowski
vacuum or to a collapsing universe with a negative cosmological
constant. Once this happens, the corresponding part of the universe
effectively disappears form consideration, as if it were falling to
a sink from which it never returns back. One of the results obtained
in \cite{Ceresole:2006iq} was that the  probability of a decay to a
collapsing space with a negative vacuum energy may be much greater
than the decay probability of a de Sitter space to a  Minkowski
space estimated in \cite{Kachru:2003aw}. We will briefly describe
this result in Section \ref{tunnsink}.

In Section \ref{currents} we will discuss a special role of the
incoming probability currents and the corresponding probability
charges in anthropic considerations. In Section \ref{comoving} we
will study these currents and charges in the comoving coordinates
(local description) and show that the results of our investigation coincide with the
results of the approach proposed in \cite{Bousso:2006ev}, without
any need to appeal to holography. In Section \ref{CVW} we will
describe one of the volume-weighted probability distributions
proposed in \cite{LLM,Bellido} and studied in \cite{Ceresole:2006iq}
in the context of the string landscape scenario. This distribution
is very similar to the comoving probability distribution, so we will
call it `pseudo-comoving':  it does not reward different parts of
the universe for the different speed of their expansion.  This
probability measure naturally appears when one studies the physical
volume of different parts of the universe at the hypersurface of
equal time, but measures the time in units of $H^{{-1}}$ along each
geodesic. In these units, all parts of the universe expand at the
same rate, which is why the map of the universe remains similar to
the map in the comoving coordinates. However, unlike the comoving
probability distribution, this probability distribution takes into
account the overall  growth of the  volume of the universe, and
therefore it leads to  different predictions, which are very
sensitive to the properties of the sinks in the landscape
\cite{Ceresole:2006iq}.

In Section \ref{VW} we will describe another volume-weighted
probability measure proposed in \cite{Eternal,LLM,Bellido}. We will
call this measure `standard,' because it calculates the physical
volume of different regions of inflationary universe taking into
account their expansion proportional to  $e^{H_{i}t}$, where $t$ is
measured in the standard physical units, such as the Planck time $M_{p}^{-1}$, or the string time $M_{s}^{-1}$.
Here $H_{i}$ are the Hubble constants in different dS spaces. An advantage of this probability measure is that the standard time, which measures the number of oscillations, is suitable for the description of chemical and biological processes, unlike the  time measured in units of $H^{{-1}}$, which corresponds to the logarithm of the distance between galaxies. Therefore one may argue that the standard probability measure may be better suited for anthropic purposes. The
results of the calculation of the probability currents and charges
in this case  are almost completely insensitive to the existence of
the sinks.

In Section \ref{BB} we will analyze the problem of Boltzmann brains
and show that the comoving probability distribution, which provides
a local description of the universe, and the pseudo-comoving
probability distribution, which does not reward growth, are not
entirely immune to the Boltzmann brain  problem. Meanwhile the
`standard' volume-weighted probability measure proposed in
\cite{Eternal,LLM,Bellido}  solves this problem.

One may wonder whether the solution of the BB problem may coexist with the solution of other problems, such as the cosmological constant problem. In Section \ref{CC} we will describe the anthropic solution of the cosmological constant (CC) problem in the string landscape scenario using the standard volume-weighted probability measure. We will argue there that the anthropic solution of the  CC problem in the string landscape scenario with a sufficiently large number of dS vacua may remain valid for a large class of the probability measures.

Finally, in the  Section \ref{disc} we discuss other problems of
different probability measures. We also argue there  that, despite
all of the uncertainties related to quantum cosmology, we can still
use the anthropic principle to explain many properties of our part
of the universe and impose strong constraints on particle physics
and cosmology. The only thing that we need to do is to study
conditional probabilities and use simple facts of our life as
observational data,  in  the same way as we use other observational
and experimental data in developing a picture of our world.

\section{Decay of de Sitter vacua and sinks in the landscape}\label{dS}

Before we start our discussion of  probabilities, we must remember some basic facts about the mechanism of  jumping from one vacuum to another. There are two related mechanisms to do so: due to tunneling \cite{Coleman:1980aw} and due to a stochastic diffusion process \cite{Starobinsky:1986fx,Linde:1991sk}.

Tunneling produces spherically symmetric universes. They look like growing bubbles for an outside observer, and like open homogeneous infinite universes from the inside observer. If the tunneling goes to dS space, the interior of the bubbles expands exponentially. From the point of view of an outside observer, the bubble walls continue moving with a speed approaching that of light,  but in comoving coordinates their size approaches some maximal value and freezes. The maximal value depends on the time when the bubble is formed; it is exponentially smaller for  bubbles formed later on \cite{Guth:1982pn}. If the tunneling goes to the state with a negative vacuum energy $V$, the infinite universe inside it collapses within a time of the order $|V|^{-1/2}$, in Planck units.

\begin{figure}[hbt]
\centering
\includegraphics[scale=0.25]{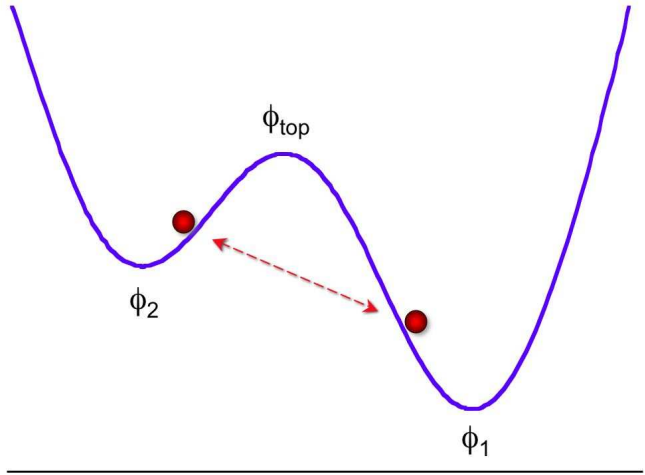} \caption{Coleman-De Luccia tunneling may go in both directions. A surprising feature of this process is that the tunneling in general occurs not from one minimum of the potential to another minimum, but form one wall of the potential to another wall. }\label{2min}
\end{figure}

Consider two dS vacua  $dS_{i}$ with the vacuum energy density $V_{i} = V(\phi_{i})$, Fig. 1. Without taking gravity into account, the tunneling may go only from the upper minimum to the lower minimum, but in the presence of gravity tunneling may occur in both directions, which is emphasized in Fig. 1. According to Coleman and De Luccia \cite{Coleman:1980aw}, the tunneling probability from $dS_{1}$ to $dS_{2}$ is given by
\begin{equation}
\label{prob} \Gamma_{12} = e^{-B} = e^{-S(\phi)+S_1},
\end{equation}
where $S(\phi)$ is the Euclidean action for the tunneling trajectory, and $S_1=S(\phi_1) $ is the Euclidean action for the initial configuration $\phi = \phi_1$,
\begin{equation}
\label{action2} S_1 = - {24\pi^2\over V_1} <0\ .
\end{equation}
This action has a simple sign-reversal relation to the entropy of de Sitter space ${\bf S_1}$:
\begin{equation}
\label{action2a} {\bf S_1} = - S_1 =+ {24\pi^2\over V_1}\ .
\end{equation}
Therefore the decay time of the metastable dS vacuum  $t_{\rm decay} \sim \Gamma^{-1}_{12}$ can be represented in the following way:
\begin{equation}
\label{decaytime} t_{\rm decay} = e^{S(\phi)+\bf S_1} = t_r \ e^{S(\phi)}\ .
\end{equation}
Here $t_{r} \sim e^{\bf S_{1}}$ is the so-called recurrence time for the vacuum $dS_{1}$.\footnote{Throughout the paper, we will assume that all decay rates are exponentially small,  so one can ignore subexponential factors in the expressions for the tunneling probabilities. Indeed, any subexponential factors can be ignored as compared to the decay rates of the type of $e^{-S_{1}}$, where $S_{1}$ is the entropy of dS state with the cosmological constant $\Lambda = 10^{-120}$. However,  in the situations where the decay rates are not too strongly suppressed, one should be careful about these factors, especially if  we measure time in units of $H_{1}^{-1} \sim 10^{60}$, where $H_{1}$ is the present value of Hubble constant.} 

Whereas the theory of tunneling developed in \cite{Coleman:1980aw} was quite general, all examples of tunneling studied there described the thin-wall approximation, where the tunneling occurs from one  minimum of the potential and proceeds directly to another minimum. This made the interpretation of the process rather simple. However, in the cases where the thin-wall approximation is not valid, the tunneling occurs not from the minimum but from the wall, which makes interpretation of this process in terms of the decay of the initial vacuum  less obvious.

The situation becomes especially confusing when the potential is very flat on the way from one minimum to another, $V'' < V$, in Planck units. In this case the Coleman-De Luccia instanton becomes replaced by the instanton describing tunneling from the top of the effective potential back to the same top of the effective potential. The corresponding instanton represents the limiting configuration of Fig. \ref{2min} when the two red balls meet at the top. According to Hawking and Moss \cite{Hawking:1981fz}, the probability of tunneling from the minimum 1 to the minimum 2 is given by
\begin{equation}
\label{HM} \Gamma_{12} = e^{-S_{\rm top}+S_1}= \exp\left(-{24\pi^2\over V(\phi_{1})}+{24\pi^2\over V(\phi_{\rm top})}\right) \ .
\end{equation}
Here $\phi_{\rm top}$ corresponds to the top of the barrier separating the two minima. Initial interpretation of this result was rather obscure because the corresponding instanton seemed to describe a homogeneous tunneling, $\phi = \phi_{\rm top}$, which does not interpolate between any minima of the potential. A homogeneous jump corresponding to this instanton would be  impossible in an infinite (or exponentially large) inflationary universe. Moreover, from the derivation of this result it was not clear why the tunneling should occur to the top of the potential instead of going directly to the second dS minimum. The situation becomes especially confusing for the case with many minima and maxima (the landscape), because  the result obtained in \cite{Hawking:1981fz} suggested that it is very easy to tunnel through high mountains if anywhere in the landscape there is a maximum with the height $V(\phi_{\rm top}) \approx V(\phi_{1})$. In fact, from the derivation it was not obvious whether the tunneling should go to the maximum instead of going directly to the next minimum, since the instantons with a constant field $\phi_{2}$ also exist. These conclusions seem obviously wrong, but why?

One of the best attempts  to clarify the situation was made by Gen and Sasaki \cite{Gen:1999gi}, who described the tunneling using Hamiltonian methods in quantum cosmology, which avoided many ambiguities of the Euclidean approach. But even their investigation did not allow us to completely resolve the paradoxes formulated above.

A proper interpretation of the Hawking--Moss tunneling was achieved only after the development of the stochastic approach to inflation \cite{Starobinsky:1986fx,Goncharov:1986ua,Linde:1991sk,LLM}. One may consider quantum fluctuations of a light scalar field $\phi$ with $m^2 = V'' \ll H^2 = V/3$.
During each time interval $\delta t = H^{-1}$ this scalar field experiences quantum jumps with the wavelength $\sim H^{-1}$ and with a typical amplitude $\delta\phi = H/2\pi$.
Then the wavelength of these fluctuations grows exponentially.
As a result, quantum fluctuations lead to a local change of the amplitude of the field $\phi$, which looks homogeneous on the horizon scale $H^{-1}$.
From the point of view of a local observer, this process looks like a Brownian motion of the homogeneous scalar field.
If the potential has a dS minimum at $\phi_1$ with $m\ll H$, then eventually the probability distribution to find the field with the value $\phi$ at a given point becomes time-independent,
\begin{equation}
\label{E38a} P(\phi) \sim \exp\left(-{24\pi^2\over V(\phi_{1})}+{24\pi^2\over V(\phi)}\right) \ .
\end{equation}
This probability distribution shows that the probability of a Brownian motion from the configuration where the horizon size domain contains the field $\phi_{1}$ to the configuration where it contains the field $\phi$ is exponentially  suppressed by a factor of $\exp\left(-{24\pi^2\over V(\phi_{1})}+{24\pi^2\over V(\phi)}\right)$. Once the scalar field climbs up to the top of the barrier, it can fall from it to the next minimum, which completes the process of ``tunneling'' in this regime. That is why the probability to gradually climb to the local maximum of the potential at $\phi = \phi_{\rm top}$ and then fall to another dS minimum is given by Hawking-Moss expression (\ref{HM}) \cite{Starobinsky:1986fx,Goncharov:1986ua,Linde:1991sk,LLM}.

The distribution $P(\phi)$, which gives the probability to find the field $\phi$ at a given point, has a simple interpretation as a fraction of the {\it comoving} volume of the universe corresponding to each of the dS vacua. Unlike the physical volume of the universe, the comoving volume does not grow when the universe expands.  To distinguish the comoving probability distribution form the volume-weighted probability distributions taking into account expansion of the universe, in \cite{Eternal,LLM} the comoving distribution was called $P_{c}$, whereas the volume-weighted probability distribution was called `physical' and denoted by $P_{p}$. Interpretation of $P_{c}$ can be understood as follows: At some initial moment one divides the universe into many domains of the same size, assigns one point to each domain, and follows the subsequent distribution  $P_{c}(\phi)$ of the points where  the scalar  field takes the value $\phi$. Physical probability distributions may differ from each other by the choice of  time  parametrization. For example, if one measures time in units of $H^{{-1}}$, different parts of the universe in these coordinates expand at the same rate. In this paper we will call the corresponding distribution `quasi-comoving,' see Section \ref{CVW}. We will call `standard' the physical probability distribution taking into account  different rates of expansion of different parts of the universe, see Section \ref{VW}.
To avoid accumulation of various indices, in this paper we will not write the indices ``c'' and ``p'' near the probability distributions, but we will specify each time what kind of distribution we are calculating.\footnote{Note that the probability distribution  $P_{p}$ introduced in \cite{Eternal,LLM} in general  is not supposed to be a global distribution describing all parts of the universe at once. Instead of that, one should consider a single causally connected domain of size $O(H^{-1})$, put there many `observers' and check what they are going to see after some time $t$. Even though these observers eventually become exponentially far from each other, formally they still belong to a causally connected part of the universe, which allows a local description.}

A necessary condition for the derivation of Eq. (\ref{E38a}) in  \cite{Starobinsky:1986fx,Goncharov:1986ua,Linde:1991sk,LLM} was the requirement that $m^2 = V'' \ll H^2 = V/3$.
This requirement is violated for all known scalar fields at the present (post-inflationary) stage of the evolution of the universe. Therefore the situation with the interpretation of the Coleman-De Luccia tunneling for $ V'' \geq   V/3$ remains somewhat unsatisfactory. In this paper we will follow the standard lore, assume that this approach is correct, and study its consequences, but one should keep this problem in mind.

Following \cite{Lee:1987qc} (see also \cite{Garriga:1997ef,Dyson:2002pf,Susskind:2003kw}), we will look for a probability distribution $P_{i}$ to find a given {\it point} in a state with the vacuum energy $V_{i}$ and will try to generalize the results for the probability distribution obtained above by  the stochastic approach to inflation.
The main idea is to consider CDL tunneling between two dS vacua, with vacuum energies $V_{1}$ and $V_{2}$, such that $V_{1} < V_{2}$, and to study the possibility of tunneling in both directions, from $V_{1}$ to $V_{2}$, or vice versa.

The action on the tunneling trajectory, $S(\phi)$, does not depend on the direction in which the tunneling occurs, but the tunneling probability does depend on it.
It is given by $e^{-S(\phi)+S_1}$ on the way up, and by $e^{-S(\phi)+{ S_2}}$ on the way down  \cite{Lee:1987qc}. Let us  assume that the universe is in a stationary state, such that the comoving volume of the parts of the universe going upwards is balanced by the comoving volume of the parts going down.
This can be expressed by the detailed balance equation
\begin{equation}
\label{balance} P_{1}\, e^{-S(\phi)+S_1} = P_{2}\, e^{-S(\phi)+S_2} \ ,
\end{equation}
which yields (compare with Eq.~(\ref{HM}))
\begin{equation}
\label{weinb} {P_{2}\over P_{1}} = e^{-S_{2}+S_1} = \exp\left(-{24\pi^2\over V_1}+{24\pi^2\over V_2}\right) \ ,
\end{equation}
independently of the tunneling action $S(\phi)$.

Equations (\ref{E38a}) and (\ref{weinb}) imply that the fraction of the comoving volume of the universe in a state $\phi$ (or $\phi_{2}$) different from the ground state $\phi_{1}$ (which is the state with the lowest, but positive, vacuum energy density) is proportional to $C_{1}\, \exp\bigl({24\pi^2\over V(\phi)}\bigr)$, with the normalization coefficient $C_{1} = \exp\bigl(-{24\pi^2\over V_{1}}\bigr)$. The probability distribution $C_{1}\, \exp\left({24\pi^2\over V(\phi)}\right)$ coincides with the square of the Hartle--Hawking wave function describing the ground state of the universe \cite{Hartle:1983ai}. It has a simple physical meaning: The universe wants to be in the ground state $\phi_{1}$ with the lowest possible value of $V(\phi)$, and the probability of deviations from the ground state is exponentially suppressed.
This probability distribution also has a nice thermodynamic interpretation in terms of dS entropy  ${\bf S}$ \cite{Linde:1998gs}:
\begin{equation}
\label{weinb2} {P_2\over P_{1}} = e^{{\bf S_{2}}-{\bf S_{1}}} = e^{{\bf \Delta S}}\ .
\end{equation}
Here, as before, ${\bf S_{i}} = - S_{i}$. This result and its thermodynamic interpretation played a substantial role in the discussion of the string theory landscape \cite{Susskind:2003kw}.

Investigation of the stationary probability distribution alone does not give us a full picture. For example, the probability distribution (\ref{E38a}) tells us about  the fraction of the comoving volume of the universe in a given state, but it tells us nothing about the evolution towards this state. A partial answer to this question can be given by investigation of the stochastic diffusion equations describing the evolution of the scalar field in the inflationary universe. But now, instead of looking for the most probable outcome of the evolution, one should follow the evolution backwards and look for the initial condition $\phi_{1}$ for the trajectories which bring the field to its final destination $\phi$. In the stationary regime considered above, the corresponding solution looks very similar to (\ref{E38a}) \cite{LLM}:
\begin{equation}
\label{E38aa} P(\phi) \sim \exp\left(-{24\pi^2\over V(\phi_{1})}+{24\pi^2\over V(\phi)}\right) \ .
\end{equation}
In this equation, however, $\phi_{1}$ is not the position of the ground state, but a position of an arbitrary initial point for the diffusion process which eventually brings us to the point $\phi$. As we see, the probability is maximized by the largest possible value of $V(\phi_{1})$. Interestingly, the expression $ \exp\bigl(-{24\pi^2\over V(\phi_{1})}\bigr)$ describing the probability of initial conditions coincides with the expression for the  square of the tunneling wave function describing creation of a closed dS universe ``from nothing'' \cite{Linde:1983mx}, whereas the second term looks like the square of the Hartle-Hawking wave function describing the ground state of the universe. In the stationary regime the squares of these two wave functions coexist in the same equation, but they provide answers to different questions.

However, this stationary distribution does not apply to the processes  during  slow-roll inflation; in order to obtain a stationary distribution during inflation one should take into account the growth of the physical volume of the universe   \cite{Eternal,LLM}.  Moreover,  this distribution does not necessarily apply to the string theory landscape either, because in the KKLT scenario there are no stable dS vacua that could serve as a ground state of the universe.
Metastability of dS space in the KKLT construction was emphasized in \cite{Kachru:2003aw} and in many subsequent papers.
Here we would like to look at this issue in a more detailed way.

\section{Tunneling to a collapsing universe with a negative vacuum energy}\label{tunnsink}

Stationarity of the probability distribution (\ref{weinb2}) was achieved because the lowest dS state did not have anywhere further  to fall.
Meanwhile, in string theory all dS states are metastable, so it is always possible for a  dS vacuum to decay \cite{Kachru:2003aw}.
It is important that if it decays by production of the bubbles of 10D Minkowski space, or by production of bubbles containing a collapsing open universe with a negative cosmological constant, the standard mechanism of returning back to the original dS state  no longer operates.\footnote{One may speculate about the possibility of quantum jumps from Minkowski space to dS space \cite{Linde:1991sk}, or even about the possibility of jumps back through the cosmological singularity inside each of the bubbles, but we will not discuss these options here.} These processes work like sinks for the flow of probability in the landscape. Because of the existence of the sinks, which are also called  terminal vacua, the fraction of the comoving volume in the dS vacua will decrease in time.

The first estimates of the probability to tunnel to the sink
made  in  \cite{Kachru:2003aw} were rather instructive and simultaneously rather optimistic. First of all, it was  shown in \cite{Kachru:2003aw} that if the decay of the metastable dS vacua occurs due to tunneling through a barrier with positive scalar potential, then the instanton action $S(\phi)$ is always negative, and therefore the decay always happens during the time shorter than the recurrence time $t_{r}$:
\be
 t_{\rm decay} = e^{S(\phi)+\bf S_1} < t_{r} = e^{\bf S_1}  \ .
 \ee
On the other hand, if the tunneling occurs, for example, from our vacuum with $V_{1} \sim 10^{-120} $ in Planck units through the barrier with a much greater $V$, or if we are talking about the Hawking-Moss tunneling to $V_{2} \gg V_{1}$, then the decay time in the first approximation would coincide with the recurrence time, i.e. our vacuum  would be incredibly  stable: $ t_{\rm decay} \sim e^{24\pi^{2}\over V_{1} }\sim 10^{10^{120}}$ years.

This result can be directly applied to the simplest KKLT model where the tunneling occurs through the positive barrier separating the metastable dS vacuum and the supersymmetric 10D vacuum.
However, the situation with the tunneling to AdS vacua for a while  remained much less clear  \cite{Banks:2005ru,Aguirre:2006ap,Bousso:2006am} because it could involve tunneling through the barriers with  $V< 0$.

This problem was recently analyzed in Ref. \cite{Ceresole:2006iq}. In that paper, we found many BPS domain wall solutions separating  different AdS vacua in the landscape. This can be done at the first stage of the landscape construction, prior to the uplifting, when  one finds all stable supersymmetric AdS vacua of the theory. Supersymmetry allows these vacua to coexist without expanding and ``eating'' each other. In all cases when the superpotential does not vanish across the domain wall, the domain wall solutions separating different vacua can be represented as the walls of the CDL bubbles of  infinitely large size \cite{Cvetic:1996vr,Ceresole:2006iq}.
For such bubbles, the tunneling action is  infinitely large, and the vacuum decay is impossible. This fact is related to the supersymmetry of the different vacua \cite{Weinberg:1982id}, and of the interpolating BPS wall solutions.

However, after the uplifting, which is required to obtain dS minima in the KKLT construction \cite{Kachru:2003aw},  supersymmetry becomes broken.
For example, in the simplest KKLT-based models the  gravitino mass squared in our vacuum is directly related to the required amount of uplifting, which  almost exactly coincides with the depth of the initial AdS vacuum prior to the uplifting: $m^{2}_{3/2} \approx |V_{\rm AdS}|/3$ \cite{Kallosh:2004yh}.
If we perform the uplifting in the theory with many different AdS minima, then only some of them will be uplifted high enough to become dS minima. Supersymmetry no longer protects them from decaying to the lower vacua. This may lead to a relatively rapid decay of the uplifted dS vacuum due to the creation of  bubbles describing collapsing open universes with a negative vacuum energy density. For brevity, we will sometimes call this process the decay to AdS vacua, but one should remember that in reality we are talking about tunneling to a collapsing space.  According to \cite{Ceresole:2006iq},  the typical decay rate for this process can be estimated as
$\Gamma \sim \exp{C M_{p}^{2}\over m^{2}_{3/2}}$. For the gravitino mass in the 1 TeV range one finds suppression in the range of $\Gamma \sim 10^{-{10^{34}}}$  \cite{Ceresole:2006iq}, which is much greater than the expected rate of the decay to Minkowski vacuum, or to a higher dS vacuum, which is typically suppressed by the factors such as $10^{-10^{120}}$. For superheavy gravitinos, which do appear in certain versions of the KKLT construction, vacuum decay rates can be even higher \cite{Page:2006dt}, which may lead to an anthropic upper bound on the degree of supersymmetry breaking in string theory.\footnote{I am grateful to Steve Shenker for the discussion of this issue.} Other possible decay channels for the uplifted dS space were discussed in \cite{Frey:2003dm,Green:2006nv,Danielsson:2006jg}.

The fact that the  decay to the collapsing AdS space can be so probable may lead to considerable changes to the standard picture of the landscape of dS vacua in thermal equilibrium. We are going to discuss this question now.

\section{Currents in the landscape with sinks}\label{currents}

\begin{figure}[hbt]
\centering
\includegraphics[scale=0.23]{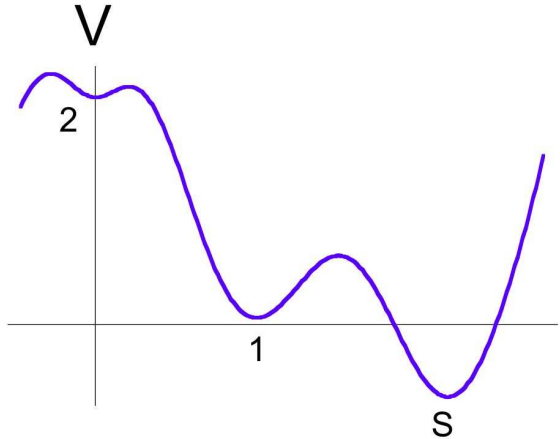} \caption{A potential with two dS minima and a sink.}\label{1sink}
\end{figure}

To make our study as simple as possible, we will begin with an investigation of a simple model describing two dS minima and one AdS minimum, denoted by 1, 2, and S in Fig. \ref{1sink}.

We will begin with the investigation of this process in comoving coordinates, i.e. ignoring the expansion of the universe. To get a visual understanding of the process of bubble formation in comoving coordinates, one may paint black all of its parts corresponding to one of the two dS states, and paint white the parts in the other dS state. Then, in the absence of  sinks in the landscape, the whole universe will become populated by white and black bubbles of all possible sizes. Asymptotically, the universe will approach a stationary regime;  the whole universe on average will become gray, and the level of gray asymptotically will remain constant.

Suppose now that some parts of the universe may tunnel to a state with a negative cosmological constant.
These parts will collapse, so they will not return to the initial dS vacua.
If we paint such parts red, then the universe, instead of reaching a constant shade of gray, eventually will look completely red. This is what we would find if we studied the properties of the universe at any given {\it point}. The probability to find the universe in a given state at a given point is given by  the comoving probability distribution $P_{i}$.

To describe this process, instead of the detailed balance equation (\ref{balance}) one should use the  ``vacuum dynamics'' equations \cite{Garriga:2005av,Ceresole:2006iq}:
\begin{equation}
\label{v01} \dot P_{1} = - J_{1s} -J_{12} + J_{21} \ ,
\end{equation}
\begin{equation}
\label{v02} \dot P_{2} = - J_{2s} -J_{21}+J_{12} \ .
\end{equation}
Here $J_{ij} = P_{j}\, \Gamma_{ji}$, where $ \Gamma_{ji}$ is the decay rate if the $j$ vacuum with respect to the bubble formation of the vacuum $i$. In particular, $ J_{1s} = P_{1}\,e^{-C_{1}}$ is the probability current from the lower dS vacuum to the sink,  i.e. to a collapsing universe, or to a Minkowski vacuum, $ J_{2s} = P_{2}\, e^{-C_{2}}$  is the probability current from the upper dS vacuum to the sink, $ J_{12} =P_{1}\, e^{-{\bf S_{1}}+|S(\phi)|}$ is the probability current from the lower dS vacuum to the upper dS vacuum, and $ J_{21} = P_{2}\, e^{-{\bf S_{2}}+|S(\phi)|}$ is the probability current from the upper dS vacuum to the lower dS vacuum. Combining this all together, gives us the following set of equations for the probability distributions:
\begin{equation}
\label{v1} \dot P_{1} =  -P_{1}\, (\Gamma_{1s} +\Gamma_{12}) +  P_{2}\,\Gamma_{21} \ ,
\end{equation}
\begin{equation}
\label{v2} \dot P_{2} =  -P_{2}\, (\Gamma_{2s} +\Gamma_{21}) +  P_{1}\,\Gamma_{12} \ .
\end{equation}
(We ignore here possible sub-exponential corrections, which appear, e.g., due to the difference in the initial size of the bubbles etc.)

The distributions $P_{i}$ play the role of the accumulated charges of the probability currents. We will also introduce equations for the charges for the {\it  incoming }probability currents $J_{12}$ and $J_{21}$:
\begin{equation}
\label{v1b} \dot Q_{1} =  J_{21} = P_{2}\,\Gamma_{21} \ ,
\end{equation}
\begin{equation}
\label{v2b} \dot Q_{2} =  J_{12}= P_{1}\,\Gamma_{12} \ .
\end{equation}
These charges take into account only the incoming probability flux, ignoring the outgoing currents.

Before solving these equations in various regimes, let us discuss the physical interpretation of the functions $P_{i}$ and $Q_{i}$.

The function $P_{i}$ describes the probability to find a given {\it point} in a particular state (in a particular dS vacuum or in  a state with a particular field $\phi$). Equivalently, it describes the fraction of {\it comoving volume} of the universe in a particular state, or the fraction of a proper time spent by a given point in this state. This function can be useful if one wants to get a map of the multiverse.

However, when the bubbles of a new phase expand, their interior eventually becomes an empty dS space devoid of any observers. If we are  usual observers born after reheating of the inflationary universe, then one may argue that the probability to be born in the bubble $dS_{i}$ is proportional not to the volume distribution $P_{i}$, but to the frequency of the new bubble production, which is related to the sum of all {\it incoming} probability currents $\dot Q_{i} = \sum_{j} J_{ij}$. 

A closely related fact was emphasized a long time ago, in the paper where we performed the first detailed investigation of the probability distribution to live in a continuous set of vacua with different properties  \cite{Bellido}.  The main idea was to find all parts of the universe at the hypersurface of the end of inflation, or at the hypersurface of a given  temperature at a given time after the beginning of inflation. After that, one should compare the relative volumes of different parts with these properties containing different values of those fields or parameters which we would like to determine using the anthropic considerations. The way to achieve this goal, which was proposed in  \cite{Bellido}, was to calculate the incoming probability currents through the hypersurface of the end of inflation.  (In \cite{Bellido} the incoming probability current at the hypersurface of the end of inflation was denoted by ${\cal P}$, to distinguish it from the probability distribution $P$ studied in \cite{LLM}. Here we use a different system of notations.)

A new feature of the string landscape scenario is that each geodesic may enter a vacuum of the same type, or the hypersurface of the end of inflation, many times, when the bubbles of the new phase are produced over and over again, and life reemerges there. Each of these entries should be counted separately when calculating the probability of the emergence of life. The integrated probability current in this context was introduced in \cite{Garriga:1997ef}.

Starting from this point, one can use  several different methods for the calculation of probabilities, depending on various assumptions.

\section{Comoving probabilities and incoming currents}\label{comoving}

The simplest possible probability measure appears if one argues that when we are trying to explain the properties of our world as {\it we} see it, we should not care about other observers. Instead we should concentrate on our own history. Because of the possible quantum jumps, our worldline could wonder many times between different dS states. Then one may argue that the probability for any given observer to find himself in a $dS_{i}$ state is proportional to the probability that his worldline entered this vacuum. But this is the definition of the charges $Q_{i}$, which are given by the integrated {\it incoming } probability currents.



One should take into account each such entry (or re-entry), and multiply the total number of such entries by the probability that each entry leads to the emergence of life as we know it \cite{Bousso:2006ev}. The last part (which we will not consider in this paper) implies, in particular, that we should pay special attention to the bubbles having  inflationary universes inside, since otherwise the bubbles will be empty open universes unsuitable for life \cite{Freivogel:2005vv}.

At the first glance, it may seem very difficult to obtain  $Q_{i}$ using our system of differential equations. Fortunately, the corresponding procedure is quite straightforward if one uses the method of integration of these equations along the lines of \cite{Starobinsky:1986fx,Garriga:2005av}.

Indeed, let us write integrated equations (\ref{v1}), (\ref{v2}), (\ref{v1b}), (\ref{v2b}) in terms of the integrals ${\cal P}_{i} = \int_{0}^{\infty} P_{i}dt$:
\begin{equation}
\label{v11} P_{1}(\infty) -P_{1}(0) =  -{\cal P}_{1}\, (\Gamma_{1s} +\Gamma_{12}) +  {\cal P}_{2}\,\Gamma_{21} \ ,
\end{equation}
\begin{equation}
\label{v21} P_{2}(\infty)-P_{2}(0) =  -{\cal P}_{2}\, (\Gamma_{2s} +\Gamma_{21}) +  {\cal P}_{1}\,\Gamma_{12} \ ,
\end{equation}
\begin{equation}
\label{v11b}  Q_{1}(\infty) = {\cal P}_{2}\,\Gamma_{21} \ ,
\end{equation}
\begin{equation}
\label{v21b} Q_{2}(\infty) =  {\cal P}_{1}\,\Gamma_{12} \ .
\end{equation}
We will be interested in the investigation of  systems with sinks, in which case $P_{1}(\infty) = 0$ \cite{Ceresole:2006iq}. Assume for definiteness that  $P_{1}(0) = 0$, and $P_{2}= 1$, i.e. we consider the system which initially was in its upper dS vacuum.
In this case the system of equations above gives
\begin{equation}
\label{v21c}q_{21}=  {Q_{2}(\infty)\over Q_{1}(\infty)} =  {\Gamma_{12}\over  \Gamma_{1s} +\Gamma_{12}} .
\end{equation}
On the other hand, if initially the system was in the lower dS vacuum, $P_{1}(0) = 1$, $P_{2}= 0$, then the same equations give
\begin{equation}
\label{v21cqq}q_{21} = {Q_{2}(\infty)\over Q_{1}(\infty)} =  1 .
\end{equation}
If one remembers that   the relative probability that the lower vacuum  jumps up   is $\epsilon =  {\Gamma_{12}\over  \Gamma_{1s} +\Gamma_{12}} $, and the relative probability to jump to  the sink is $1-\epsilon =  {\Gamma_{1s}\over  \Gamma_{1s} +\Gamma_{12}} $, then one finds that our results are equivalent to the results obtained by Bousso
\cite{Bousso:2006ev} by a different method. 

Our method of calculation of comoving probabilities does not require any reference to  holography,\footnote{This is not surprising since inflation by its nature is opposite to the basic idea of the holographic approach. Indeed, the main advantage of inflation is its ability to erase all memory of initial and boundary conditions, which is opposite to the original idea of the holographic principle, which suggests that  all 3D gravitational dynamics can be described in terms of  the dynamics on some 2D hypersurface \cite{hologr}. In the cosmological context, this idea was gradually reduced to the derivation of various bounds on entropy  \cite{Bousso:2002ju}. One may still try to  combine eternal inflation with the original version of the holographic principle, but it is a rather challenging task \cite{Freivogel:2006xu}.} and it can be easily compared to other methods using standard terminology of eternal inflation. We are talking here about the total charges corresponding to the incoming probability currents in the comoving coordinates \cite{Bellido}, but we are applying this methodology to the  situation with many discrete dS vacua  \cite{Garriga:1997ef,Bousso:2006ev}. Quantities like that are invariant with respect to  different choices of the time variable \cite{Garriga:2005av}. They do not require introduction of any artificial cutoffs;  an exponential cutoff is naturally present here because of the existence of the sinks in the landscape.

This approach is quite interesting and informative, but it is somewhat incomplete, because it makes predictions only after we specify initial conditions for inflation. This returns us to the question of the measure of initial conditions, and to the 20 years old debate about the Hartle-Hawking wave function versus the tunneling wave function. We will discuss this question in Section \ref{BB}. Meanwhile one of the  main advantages of eternal inflation is that it makes everything that happens in an inflating universe independent of the initial conditions. That is why most of the efforts for finding the probability measure in eternal inflation were based on the global, volume-weighted probability distributions, which do not depend on initial conditions.

The dependence on the initial conditions does not automatically disqualify the local approach. In fact, different versions of this method have been used in the past for making cosmological predictions. From my perspective, the main problem with the comoving probability distribution is not the dependence on initial conditions, but the fact that, by construction, this method is not very convenient for investigation of the large scale structure of an eternally inflating universe.  We will describe now the simplest volume-weighted probability distribution, which, at the first glance, is almost indistinguishable from the comoving distribution, but which leads to different predictions.

\section{Pseudo-comoving volume-weighted measure}\label{CVW}

One of the  main advantages of inflation is that it can explain the enormously large size of the universe. Eternal inflation does even more. It can take two causally-connected regions and then make the distance between them indefinitely large. Thus a single causally connected region of the universe eventually will contain indefinitely many observers like us. If we are typical, we should study the distribution of all observers over the whole universe, and then find where most of them live. This idea and the methods of calculating probability distributions to find observers in different parts of the universe in the context of eternal inflation were developed in \cite{LLM,Bellido,Mediocr}.\footnote{This idea sometimes is called  `the principle of mediocrity.' I prefer the standard name `anthropic principle,'  because I believe that the word `anthropic' is absolutely essential in describing a proper way to calculate conditional probabilities under the obvious condition that {\it we}, rather than some abstract information-processing devices, are making the observations that we are trying to explain; see a discussion of this issue in Section \ref{disc}.}
 But one cannot find out what is typical and what is not by concentrating on a single observer and ignoring the main fraction of the volume of the universe.
Indeed, each particular observer within a finite time will die in the collapsing universe. However,  the universe eternally rejuvenates due to the exponential expansion of its various dS parts,  the total volume of the universe in different dS states continues to grow exponentially, and so does the total number of observers living there \cite{linde1982,Guth:1980zm,Steinhardt,Vilenkin:1983xq}. This process of eternal creation of new points and new observers is completely missed by the investigation of the comoving probability currents performed in the previous section.

This problem can be cured by a   tiny modification of our previous approach without changing any of our equations \cite{LLM,Bellido,Starobinsky:1994bd,Garriga:2005av,Ceresole:2006iq}. Indeed, the picture of the universe in the comoving coordinates will not change if we  study the growth of the volume of the universe, but use the units of time adjusted for the local value of the Hubble constant: $\Delta t = H^{{-1}}$. In these coordinates all parts of the universe will expand at the same
rate: During the time $\Delta t = H^{{-1}}$ all sizes grow $e$ times, the total volume will grow $e^{3}$ times, but the distribution of while, black and red bubbles will not change; it will  only be scaled by a factor   $e$ in all directions. We will call this picture `pseudo-comoving.' Using the time  as measured in units of $\Delta t = H^{{-1}}$  is equivalent to measuring time in units of the logarithm of the expansion of the universe, e.g. in the units of the logarithm of the distance between galaxies \cite{Starobinsky:1986fx,LLM,Bellido}.

The functions $P_{i}$ will depend on the expansion of the universe, but their ratios, i.e. the fraction of volume in the states $dS_{i}$,  will remain the same as in the comoving coordinates. The main thing that changes is our interpretation of the whole picture. Now we should remember that even though the whole picture on average becomes red, the total number of observers in white and black areas  continues growing exponentially. Therefore the main contribution to the charges $Q_{i}$ taking into account the exponential growth of the universe will be determined  by the integration of the probability currents in the distant future.  In such a situation, the measure of the relative probability to be born in a vacuum $dS_{i}$, which is determined by the integral of the incoming probability currents $\dot Q_{i}$ until some time cut-off, will not depend on initial conditions and will be given  by the constant ratio of the incoming currents ${\dot Q_{i}\over \sum_{j} \dot Q_{j}}$.

To take into account the exponential expansion of the universe on the formal level, one should write an extended version of the equations
(\ref{v1}), (\ref{v2}) by adding there the terms $3P_{1}$ and $3P_{2}$ describing the growth of volume (of points) due to the exponential expansion of the universe. Note that these terms do not contain $H$ because we decided, in this section, to measure time locally, in units of $H^{{-1}}$, to keep the picture similar to the one in comoving coordinates, up to an overall rescaling. As before, we are assuming that all decay rates are exponentially small, and we can write  these equations ignoring all subexponential coefficients:
\begin{equation}
\label{v1cvw} \dot P_{1} =  -P_{1}\, (\Gamma_{1s} +\Gamma_{12}) +  P_{2}\,\Gamma_{21}  + 3 P_{1}\ ,
\end{equation}
\begin{equation}
\label{v2cvw} \dot P_{2} =  -P_{2}\, (\Gamma_{2s} +\Gamma_{21}) +  P_{1}\,\Gamma_{12} + 3 P_{2}\ ,
\end{equation}
\begin{equation}
\label{v1bcvw} \dot Q_{1} =    P_{2}\,\Gamma_{21} \ ,
\end{equation}
\begin{equation}
\label{v2bcvw} \dot Q_{2} =   P_{1}\,\Gamma_{12} \ .
\end{equation}

We can also write these equations in an expanded form:
\begin{equation}
\label{v1a} \dot P_{1} =   -P_{1}\, e^{-C_{1}} -P_{1}\, e^{-{\bf S_{1}}+|S(\phi)|} + P_{2}\, e^{-{\bf S_{2}}+|S(\phi)|}+ 3 P_{1} \ ,
\end{equation}
\begin{equation}
\label{v2a} \dot P_{2} = - P_{2}\, e^{-C_{2}} -P_{2}\, e^{-{\bf S_{2}}+|S(\phi)|} + P_{1}\, e^{-{\bf S_{1}}+|S(\phi)|}+ 3 P_{2}\ ,
\end{equation}
\begin{equation}
\label{v1bcvwa} \dot Q_{1} =     P_{2}\, e^{-{\bf S_{2}}+|S(\phi)|} \ ,
\end{equation}
\begin{equation}
\label{v2bcvwa} \dot Q_{2} =   P_{1}\, e^{-{\bf S_{1}}+|S(\phi)|} \ .
\end{equation}

To analyze different  solutions of these equations, let us try to understand the relations between their parameters.
Since the entropy is inversely proportional to the energy density, the entropy of the lower level is higher, ${\bf S_{1}} > {\bf S_{2}}$. Since the tunneling is exponentially suppressed,  we have ${\bf S_{2}} > |S(\phi)|$, so we have a
hierarchy $ {\bf S_{1}} > {\bf S_{2}} > |S(\phi)|$, and therefore    $\Gamma_{12}\ll \Gamma_{21}\ll 1$. We will often associate the lower vacuum with our present vacuum state, with $S_{1} \sim 10^{{120}}$.

For simplicity, we will study here the  possibility that only the lower vacuum can tunnel to the sink, $\Gamma_{2s}= 0$,  i.e.
we will take the limit $C_{2}\to \infty$ and drop the term $- J_{2s}= - P_{2}\, e^{-C_{2}}$ in Eq. (\ref{v2}).
On the other hand, we will keep in mind the results of the previous Section, where we have found that typically the probability of the decay of a  metastable dS vacuum to a sink  can be quite high, $\Gamma_{1s}= e^{-C_{1}} \sim \exp \bigl(-{O(m_{3/2}^{-2})}\bigr) \gg e^{-{\bf S_{1}}} \sim e^{-10^{120}}$.  Therefore we  expect that ${\bf S_{1}}\gg C_{1}$.


By solving equations (\ref{v1cvw}), (\ref{v2cvw}), one can show that   the ratio $P_{2}(t)/P_{1}(t)$ approaches a stationary regime $P_{2}(t)/P_{1}(t) = p_{21} = const$.
In order to find $p_{21}$, one can add to each other our equations (without the term $ - P_{2}\, \Gamma_{2s}$, which we assumed equal to zero). This yields
\begin{equation}
\label{v3} (1+p_{21})\dot P_{1} = 3P_{1}(1+p_{21})    -P_{1}\Gamma_{1s} \ .
\end{equation}
The solution is
\begin{equation}\label{exgr}
P_{1} = {P_{2}\over p_{21}} =   \tilde P_{1}\, e^{3t} \exp\left(-{\Gamma_{1s}\over 1+p_{21}}\, t\right) \ .
\end{equation}
Here $\tilde P_{1}$ is some constant, which is equal to $P_{1}(t=0)$ if the asymptotic regime is already established at $t = 0$. The factor $ \tilde P_{1}\, e^{3t}$ shows that the overall volume grows exponentially, whereas the factor $\exp\left(-{\Gamma_{1s}\over 1+p_{21}}\, t\right)$ shows that the {\it relative} fraction of the volume in dS vacua is decreasing exponentially due to the decay to the sink    \cite{Ceresole:2006iq}.

It is most important that the {\it total} volume of space in dS vacua (and the total number of observers living there) continues growing exponentially, as $\exp\left(3-{\Gamma_{1s}\over 1+p_{21}}\, t\right)$. This fact cannot be seen in the investigation in the comoving coordinates performed in the previous section.  The factor $3 -{\Gamma_{1s}\over 1+p_{21}}$ is the fractal dimension of the domains $P_{i}$ (the same for both types of domains), see \cite{Aryal:1987vn,LLM}.

For  the (asymptotically) constant ratio  $p_{21}  = P_{2}(t)/P_{1}(t)$, from Eqs.~(\ref{v1cvw}), (\ref{v2cvw}) one finds
\begin{equation}
\label{ttt} p_{21}^{2}\Gamma_{21} -p_{21}(\Gamma_{1s} + \Gamma_{12}-\Gamma_{21}) -\Gamma_{12} = 0\ .
\end{equation}
Note that the constant $3$ disappears from this equation: The terms $3P_{i}$ only changes the overall normalization of our solutions, and drop  out from the expression for the ratio   $p_{21}  = P_{2}(t)/P_{1}(t)$. That is why  these terms have not been added explicitly to the equations in \cite{Ceresole:2006iq}.

One may consider two interesting regimes, providing two very different types of solutions. Suppose first that  $\Gamma_{1s} \ll \Gamma_{21}$\, ($ e^{-C_{1}} \ll e^{-{\bf S_{2}}+|S(\phi)|}$), i.e. the probability to fall to the sink from the lower vacuum is smaller than the probability of the decay of the upper vacuum.
In this case one recovers the previous result, Eq.~(\ref{weinb2}), which is related to the square of the Hartle-Hawking wave function, or to the thermal equilibrium between the two dS vacua:
\begin{equation}
\label{HHeq} p_{21} = {P_{2}\over P_{1}} = {\Gamma_{12}\over \Gamma_{21}} = e^{{\bf S_{2}}-{\bf S_{1}}} \ll 1 \ .
\end{equation}
It is interesting that this thermal equilibrium is maintained even in the presence of a sink if $\Gamma_{1s} \ll \Gamma_{21}$. Note that the required condition for thermal equilibrium is not  $\Gamma_{1s} \ll \Gamma_{12}$, as one could naively expect, but rather $\Gamma_{1s} \ll \Gamma_{21}$. We will call such sinks narrow.

Now let us consider the opposite regime, and assume that the decay rate of the uplifted dS vacuum to the sink is relatively large, $\Gamma_{1s} \gg \Gamma_{21}$\, ($ e^{-C_{1}} \gg e^{-{\bf S_{2}}+|S(\phi)|}$),   which automatically means that $\Gamma_{1s} \gg \Gamma_{12}$\,  ($ e^{-C_{1}} \gg e^{-{\bf S_{1}}+|S(\phi)|}$). In this ``wide sink'' regime  the solution of Eq.~(\ref{ttt}) is
\begin{equation}\label{ooo}
p_{21} = {P_{2}\over P_{1}} =  {\Gamma_{1s}\over \Gamma_{21}} = e^{{\bf S_{2}}-|S(\phi)| -C_{1}} \approx e^{{\bf S_{2}}-|S(\phi)|} \gg 1 \ ,
\end{equation}
i.e. one has  {\it an inverted probability distribution}. This result has a simple interpretation: if the ``thermal exchange'' between the two dS vacua occurs very slowly as compared to the rate of the decay of the lower dS vacuum, then the main fraction of the volume of the dS vacua will be in the state with the higher energy density, because everything that flows to the lower level rapidly falls to the sink.

Now we should remember that an important quantity to calculate for  anthropic applications is not  $p_{21}= P_{2}/P_{2}$ but $q_{21} = Q_{2}/Q_{1}$. In the previous section this quantity was calculated by integrating our equations of motion, and the results were dependent on initial conditions. In the present case, new parts of the universe (and new observers) appear exponentially faster than the old parts tunnel to the sink and die, see Eq. (\ref{exgr}). Therefore the main part of the probability current flows to dS vacua at asymptotically large values of time, and the ratio $ Q_{2}/Q_{1}$ becomes equivalent to the asymptotic ratio of the probability currents,
\be
q_{21} = {Q_{2}\over Q_{1}} =  {\dot Q_{2}\over \dot Q_{1}} = {J_{12}\over J_{21} }\ .
\ee

 In the absence of the sink,  the fraction of the comoving volume which flows to the lower dS vacuum due to the tunneling from the upper dS vacuum is equal to the fraction of the volume jumping upwards from the lowest vacuum to the higher vacuum. In other words, the two probability currents are exactly equal to each other,
\be
q_{21} =  {J_{12}\over J_{21} } = 1  \ ,
\ee
which is the essence of the detailed balance equation (\ref{balance}). Our results imply that this regime remains approximately valid even in the presence of the sink, under the condition $\Gamma_{1s} \ll \Gamma_{21}$.

On the other hand, in the regime described by Eq.  (\ref{ooo}), which occurs if the decay rate to the sink is large enough, $\Gamma_{1s} \gg \Gamma_{21}$, one has a completely different result:
 \begin{equation}
\label{vvvv} q_{21} = {J_{12}\over J_{21}} =  {P_{1}\, e^{-{\bf S_{1}}+|S(\phi)|}\over  P_{2}\, e^{-{\bf S_{2}}+|S(\phi)|}} =  e^{-{\bf S_{1}}+|S(\phi)| +C_{1}} \approx e^{-{\bf S_{1}}} \sim e^{-10^{120}} \ .
\end{equation}
Thus we have a crucial regime change at the moment when the decay rate of the lower vacuum to the sink starts competing with the decay rate of the upper dS vacuum, i.e. at the moment that we  go  from the narrow sink regime to the wide sink regime.

\section{`Standard' volume-weighted distribution: rewarding the leaders}\label{VW}

Until now, we were working in the comoving coordinates, in Section \ref{comoving}, or in the coordinates obtained from the comoving ones by a trivial scaling, in Section \ref{CVW}. This was the most conservative approach which did not reward any parts of the universe for their inflationary growth. From the point of view of inflationary cosmology, this approach may seem rather artificial, but we followed it because we wanted to compare the results of different approaches to each other, and to outline possible resolutions of some of the recently formulated paradoxes.

Now we are going to make one more step and study the volume-weighted probability distribution introduced in \cite{Eternal,LLM,Bellido}, where we measure time in the standard (e.g. Planckian) units, and take into account that the physical volume  of the universe in a $dS_{i}$ state on a hypersurface of a given time $t$ grows as $e^{3H_{i}t}$, where $H_{i}^{2} = V_{i}/3$, in the units $M_{p} = 1$. For definiteness,  we will call the resulting volume-weighted probability distribution `standard.' It may seem disappointing that the final results of the investigation should depend on the choice of the time slicing  \cite{LLM,Bellido}. One the other hand, one may argue that it is most natural to measure time in the standard units $M_{p}^{{-1}}$,  or the string time $M_{s}^{-1}$, because all local processes, oscillations,  decay rates, and the rate of the chemical and biological evolution are most naturally defined using this time variable,  instead of being controlled by the distance between galaxies, which was the essence of the time variable studied in the previous section.  

In this case, our system of equations becomes
\begin{equation}
\label{v1vw} \dot P_{1} =  -P_{1}\, (\Gamma_{1s} +\Gamma_{12}) +  P_{2}\,\Gamma_{21} +3H_{1}P_{1}\ ,
\end{equation}
\begin{equation}
\label{v2vw} \dot P_{2} =  -P_{2}\, (\Gamma_{2s} +\Gamma_{21}) +  P_{1}\,\Gamma_{12} +3H_{2}P_{2}\ ,
\end{equation}
\begin{equation}
\label{v1bvw} \dot Q_{1} =  J_{21} = P_{2}\,\Gamma_{21} \ ,
\end{equation}
\begin{equation}
\label{v2bvw} \dot Q_{2} =  J_{12}= P_{1}\,\Gamma_{12} \ .
\end{equation}
Note that the changes occur only in the upper two equations.

Using the same methods as in the previous section, one can find that
\begin{equation}
\label{v3aa} (1+p_{21})\dot P_{1} = P_{1}\, (3H_{1}+3p_{21}H_{2} -\Gamma_{1s} -p_{21}\Gamma_{2s} ) \ .
\end{equation}
which yields
\begin{equation}\label{exgrrr0}
P_{1} = {P_{2}\over p_{21}} =   \tilde P_{1}\ \exp\left({3H_{1}+3p_{21}H_{2} -\Gamma_{1s}-p_{21}\Gamma_{2s} \over 1+p_{21}}\, t\right) \ .
\end{equation}

For  the (asymptotically) constant ratio  $p_{21}  = P_{2}(t)/P_{1}(t)$, from Eqs.~(\ref{v1vw}), (\ref{v2vw}) one finds
\begin{equation}
\label{tttaa} p_{21}^{2}\Gamma_{21} -p_{21}\Bigl(3(H_{2}-H_{1}) +(\Gamma_{1s} + \Gamma_{12}) -(\Gamma_{2s}+\Gamma_{21})\Bigr) -\Gamma_{12} = 0\ .
\end{equation}
To analyze this equation, we will assume that $H_{1}$,  $H_{2}$, and their difference, $H_{2}  -  H_{1}$, are  much greater than the typical decay rates. This is indeed the case even for the present extremely small Hubble constant, $H \sim 10^{-60}$, as compared to the typical numbers encountered in our calculations for the decay rate, such as $10^{-10^{30}}$, or $10^{-10^{120}}$. We will also take into account that $\Gamma_{21} \gg \Gamma_{12}$. In this case our equation has a simple solution
\begin{equation}
\label{ttta} p_{21} = {P_{2}\over P_{1}} = {3(H_{2}-H_{1})\over \Gamma_{21}} \gg 1\ ,
\end{equation}
and we find the final expressions for $P_{i}$:
\begin{equation}\label{exgrrr}
P_{1}  =   \tilde P_{1}\ e^{3H_{2}\,t}\,  \ ,
\end{equation}\begin{equation}\label{exgrrra}
P_{2}  =   \tilde P_{1}\   {3(H_{2}-H_{1})\over \Gamma_{21}}\  e^{3H_{2}\,t}\, \ .
\end{equation}
Finally, let us calculate the ratio of the incoming probability currents, which may be important for anthropic applications:
\be\label{onesdown}
q_{12} = {Q_{1}\over Q_{2}} =  {\dot Q_{1}\over \dot Q_{2}} = {P_{2}\,\Gamma_{21}\over P_{1}\,\Gamma_{12} } =  {3(H_{2}-H_{1})\over \Gamma_{12}} \gg 1 \ .
\ee
Note that {\it the rate of decay to the sink plays no role in these results}. Let us try to understand these results as it is going to help us to analyze more complicated situations.

First of all, the volume of {\it all}\, dS vacua grows at the same rate, which practically coincides with the rate of growth of the upper dS vacuum. The reason is that after a brief delay, a finite part of the volume of the upper dS transforms into the volume of the lower dS. So the volume of the lower dS grows mostly not because of its own expansion, but because of the decay of the rapidly growing  upper dS. This is exactly the situation encountered in \cite{LLM} during a similar analysis of eternal slow-roll chaotic inflation.

Secondly, the volume of the upper dS is much greater than the volume of the lowed dS, by a factor of ${3(H_{2}-H_{1})\over \Gamma_{21}}$. More to the point, the ratio of the probability flux $\dot Q_{1}$ incoming to the lower dS is greater than the flux upwards $\dot Q_{1}$ by an even  greater factor ${3(H_{2}-H_{1})\over \Gamma_{12}}$. Suppose for example that $V_{1} = 10^{{-120}}$, $ H_{1} \sim 10^{-60}$ and $S_{1} \sim 10^{120}$, as in our vacuum. Suppose also that the instanton action $|S(\phi)|$  is much smaller than $S_{1} \sim 10^{120}$. Then the flux downwards is greater  than the flux upwards by the factor of $10^{10^{120}}$, i.e. $q_{21} \sim 10^{-10^{120}}$.

\begin{figure}[hbt]
\centering
\includegraphics[scale=0.25]{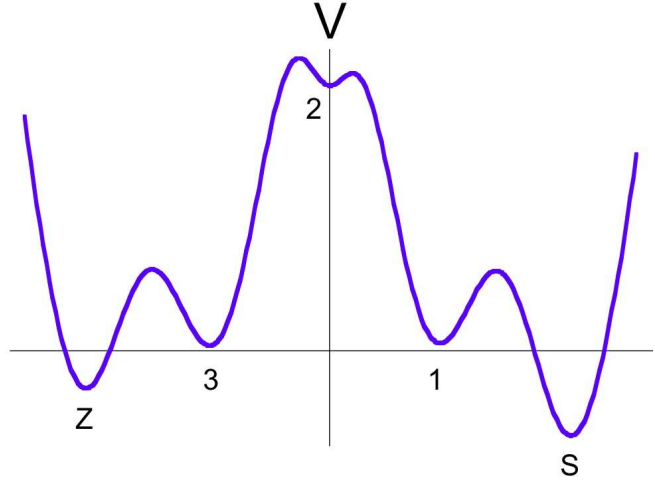} \caption{A potential with three dS minima and two sinks.}\label{2sinksnobrains}
\end{figure}

As a more complicated example, let us consider the potential with three different dS minima, as shown in Fig. \ref{2sinksnobrains}. The equations for $P_{i}$ can be written as follows: 
\begin{equation}
\label{v1vw000} \dot P_{1} =  -P_{1}\, (\Gamma_{1s} +\Gamma_{12}) +  P_{2}\,\Gamma_{21} +3H_{1}P_{1}\ ,
\end{equation}
\begin{equation}
\label{v2vw000} \dot P_{2} =  -P_{2}\, (\Gamma_{2s} +\Gamma_{21}+\Gamma_{23}) +  P_{1}\,\Gamma_{12}+P_{3}\,\Gamma_{32} +3H_{2}P_{2}\ ,
\end{equation}
\begin{equation}
\label{v3vw000} \dot P_{3} =  -P_{3}\, (\Gamma_{3z} +\Gamma_{32}) +  P_{2}\,\Gamma_{23} +3H_{3}P_{3}\ .
\end{equation}

We will be interested in the case where $H_{2}$ is much greater than all other parameters in these equations. In this case $P_{2}$ obeys a simple equation
\begin{equation}
\label{v2vw000oo} \dot P_{2} = 3H_{2}P_{2}\ ,
\end{equation}
i.e. in the first approximation $P_{2}$ does not depend on $P_{1}$, $P_{3}$:
\be
P_{2} = P_{2}(0)\, e^{3H_{2}\, t} \ .
\ee
Because of the fast growth of $P_{2}$, the terms $P_{2} \Gamma_{2i}$ eventually become the leading terms in the equations for $P_{i}$ and $Q_{i}$, for all $i\not = 2$:
\begin{equation}
\label{v1vw000i} \dot P_{i} = \dot Q_{i} =  P_{2}\,\Gamma_{2i} \ ,
\end{equation}
with the solution
\begin{equation}
\label{v1vw000i2}  P_{i} =  Q_{i} =  P_{2}\,{\Gamma_{2i}\over 3H_{2}} \ ,
\end{equation}
We find that 
\be
q_{i2} = {Q_{i}\over Q_{2}} =  {3H_{2}\over \Gamma_{i2}} \gg 1 \ ,
\ee
which agrees with the previously obtained result (\ref{onesdown}) in the limit $H_{i} \ll H_{2}$. Thus, in this limit the currents to the lower minima do not affect each other, and  the ratio of the probability currents from the upper minimum to the lower minima will be given by
\be\label{tunndown}
q_{ij} = {Q_{i}\over Q_{j}} = {\dot Q_{i}\over \dot Q_{j}} = {\Gamma_{2i}\over \Gamma_{2j}} \ , 
\ee
where $i,j \not = 2$. 

In the string landscape scenario one may have many different dS minima with  the vacuum energy density of the same order of magnitude as the energy density in the highest dS state (or many regions with flat potentials where the energy density is very high and a slow-roll eternal inflation is possible). The transition rates between these parts of the landscape may not be strongly suppressed because $e^{-\bf S}$ for these parts of the universe may not be very small. In this case the combination of all of such parts of the landscape, which one may call `the highland,' will determine the average rate of growth of volume of all other parts of the universe.\footnote{We will  assume here, as many authors do, that the landscape is {\it transversable}, which means that the process of tunneling and quantum diffusion can bring us from any part of the landscape to any other part. If the landscape consists of several completely disconnected parts, then the methods discussed in this paper will apply to the transversable part of the landscape including the domain where we live; one may need to use quantum cosmology to compare various disconnected parts of the landscape. We will discuss this issue in a separate publication.} In this case one can propose the following schematic generalization of Eq. (\ref{tunndown}), describing the relative prior probability to live in  the lower  minimum $dS_{i}$ versus $dS_{j}$:
\be\label{highlanddown}
q_{ij} = {Q_{i}\over Q_{j}} = {\dot Q_{i}\over \dot Q_{j}} = {\Gamma_{hi}\over \Gamma_{hj}} \ . 
\ee
Here $\Gamma_{hi}$ is a  probability of a transition from the highland to one of the lower states,  $dS_{i}$. The transition rate should be integrated over possible all initial states in the highland and may involve a cascade of events bringing us to the $dS_{i}$ state. Note that this generalization is schematic and oversimplified; for example, some of the transitions (or some of the parts of the highland) may involve slow roll eternal inflation. In this case our methods should be complemented by the methods developed in \cite{LLM,Bellido}. Nevertheless, this way of representing the process of cascading down from the highland will be quite useful for our subsequent discussion.

We could continue this investigation and perform a similar analysis for different, more complicated probability measures, but we will leave it for a separate investigation. We believe that we already have enough weapons to defend ourselves from the invasion of Boltzmann brains.

\section{Invasion of Boltzmann brains.}\label{BB}

The history of the BB paradox goes back to the paper by   Dyson,  Kleban and  Susskind  \cite{Dyson:2002pf}; see also \cite{Albrecht:2004ke}. They  argued that dS space is a thermal system. In such a system people, planets, and galaxies   can appear  from dS space due to thermal fluctuations, without passing through the usual stage of the big bang evolution. No formal investigation of such processes was performed so far, but simple estimates made in If one takes the typical estimate of the rate of the BB production $\Gamma_{1B}\sim 10^{-10^{50}}$ \cite{Page:2006dt,Bousso:2006xc} The probability of such events will be incredibly small, but it was argued in \cite{Dyson:2002pf} that the typical time $\tau$ required for a spontaneous non-inflationary materialization of the world similar to ours is much shorter than the time required  to jump back to the dS space with higher energy density and initiate a new stage of inflation. Therefore if we consider all observers who will ever live in an eternally existing dS universe, then most of them would be created by thermal fluctuations rather then by the rare incidents of inflation. In order to explain observational data indicating that inflation did happen in the past,  we would need to assume that even the lowest dS space cannot be stable, and its  lifetime must be much shorter than the time $\tau$.

For a while, this scary picture did not attract much attention simply because it was based on a specific way of calculating probabilities in eternal inflation, which some of us did not consider natural. In addition, we thought that we still had $10^{10^{120}}$ years or so to check whether this problem was serious.

The situation changed in an interesting way after the discovery of the KKLT mechanism \cite{Kachru:2003aw}. One of the results obtained in  \cite{Kachru:2003aw} was the upper bound on the decay time of a metastable dS state: $t < t_{r} \sim e^{S_{1}}$. This result was greeted as a confirmation of the conclusions of Ref. \cite{Dyson:2002pf}, and, simultaneously, as a resolution of the paradox formulated there  \cite{Goheer:2002vf}.

However, there was a lingering thought that  the decay rate found in \cite{Kachru:2003aw} may be too small to resolve this paradox. A more detailed investigation of this problem was performed in our paper  \cite{Ceresole:2006iq}, using the rules of calculating the probabilities similar to those used in  \cite{Dyson:2002pf} (in particular, not rewarding the rapidly growing parts of the universe). We have found that the paradox formulated in  \cite{Dyson:2002pf} can be resolved in this context, but only if the lifetime of dS states is much shorter than $\tau$; for a more precise condition see Sect. \ref{cer} below. We also found that quite often the lifetime of dS states is indeed relatively short because of the fast  decays to wide sinks \cite{Ceresole:2006iq}.

A new twist of this story is related to recent papers by Don Page  \cite{Page:2006dt,Page:2006hr}. The essence of his argument can be formulated as follows. Consider a cubic kilometer (or a cubic Megaparsec) of space and count all observers living there. We will come up with some large but finite number. In the future, this part of the universe will grow exponentially.  Some parts of it will decay and die, but just as we already mentioned, the total volume of the non-decayed parts will continue growing exponentially and eventually its size will become indefinitely large. Even if the probability of spontaneous creation of an observer in the future is incredibly small, an infinite fraction of observers will live in the future because the total volume where they can materialize is infinite. Why then  did we appear after inflation instead of being created later? Why are we so {\it atypical}? One way to avoid this paradox is to assume that the universe in the future is not expanding exponentially because its decay rate is faster than the rate of the doubling of its volume. This suggests that our universe is going to die in about $10^{10}$ years. This is not unrealistic if one uses the estimate of the lifetime of the universe $t \sim \exp\Bigl({M_{p}^{2}\over m_{3/2}^{2}}\Bigr)$ given in \cite{Ceresole:2006iq} and assumes that the gravitino is superheavy.

This way of thinking is not without its own problems, as we will discuss in Sect. \ref{disc}, but let us follow it for a while. And this will  bring us to the possible demise of the Hartle-Hawking wave function \cite{Page:2006hr}, for the reasons to be discussed in Sect. \ref{bousso}.

Finally, in their recent paper \cite{Bousso:2006xc}
Bousso and Freivogel re-formulated this problem in terms of the so-called  brains, observers created from an empty dS space by quantum fluctuations. They argued that the problem formulated in \cite{Dyson:2002pf,Albrecht:2004ke,Page:2006dt,Page:2006hr} is very serious and cannot be resolved using the global description of inflation. According to \cite{Bousso:2006xc}, the only way to solve this problem is to adopt the comoving probability distribution described in \cite{Bousso:2006ev}. If correct, this would be a very powerful conclusion, which would allow us to single out a preferable definition of measure in eternal inflation.

Since we already developed a unified framework where this question can be analyzed, let us try to find out whether this is indeed the case and what is going on with the BBs, using three different ways of slicing our universe.

The only thing that we need to do is to add an equation for the probability current of creation of the  brains:
\begin{equation}
\label{qbb} \dot Q_{BB} =  J_{1B} = P_{1}\,\Gamma_{1B} \ .
\end{equation}
Here $\Gamma_{1B}$ is the rate of the BB production in the vacuum $dS_{1}$. All other equations should remain the same because BBs just appear in $dS_{1}$ and relatively rapidly disappear in $dS_{1}$, without creating new de Sitter bubbles, so all $P_{i}$ remain unaffected.\footnote{That is, of course, if BBs are not aggressive and do not participate in any experiments that could trigger a transition to a different vacuum. We presume that this is only a prerogative of honest folk like ourselves.}

\subsection{BBs and comoving probabilities}\label{bousso}

Since the main part of the problem was already solved in Sect. \ref{comoving}, we will give here only  final results. If the universe begins in the upper dS vacuum, $dS_{2}$, then we find
\begin{equation}
\label{qbb1} {Q_{BB}\over Q_{1}} =    {\Gamma_{1B}\over \Gamma_{1s}+\Gamma_{12} } \   .
\end{equation}
 We see that if we consider wide sinks with $\Gamma_{1B}\gg \Gamma_{1s}+\Gamma_{12}$,   then the comoving observer spends most of his life (or lives) as an OO (ordinary observer) rather than as a BB.

Meanwhile, if the process begins in the lower vacuum, we have
\begin{equation}
\label{qbb2} {Q_{BB}\over Q_{1}} =    {\Gamma_{1B}\over \Gamma_{12} } \   .
\end{equation}
The standard assumption of \cite{Dyson:2002pf,Page:2006dt,Page:2006hr,Bousso:2006xc} is that the probability of BB production is much higher than the probability to jump to the higher dS. This implies that the poor guy starting in the lower vacuum is doomed to being a BB.

\begin{figure}[hbt]
\centering
\includegraphics[scale=0.28]{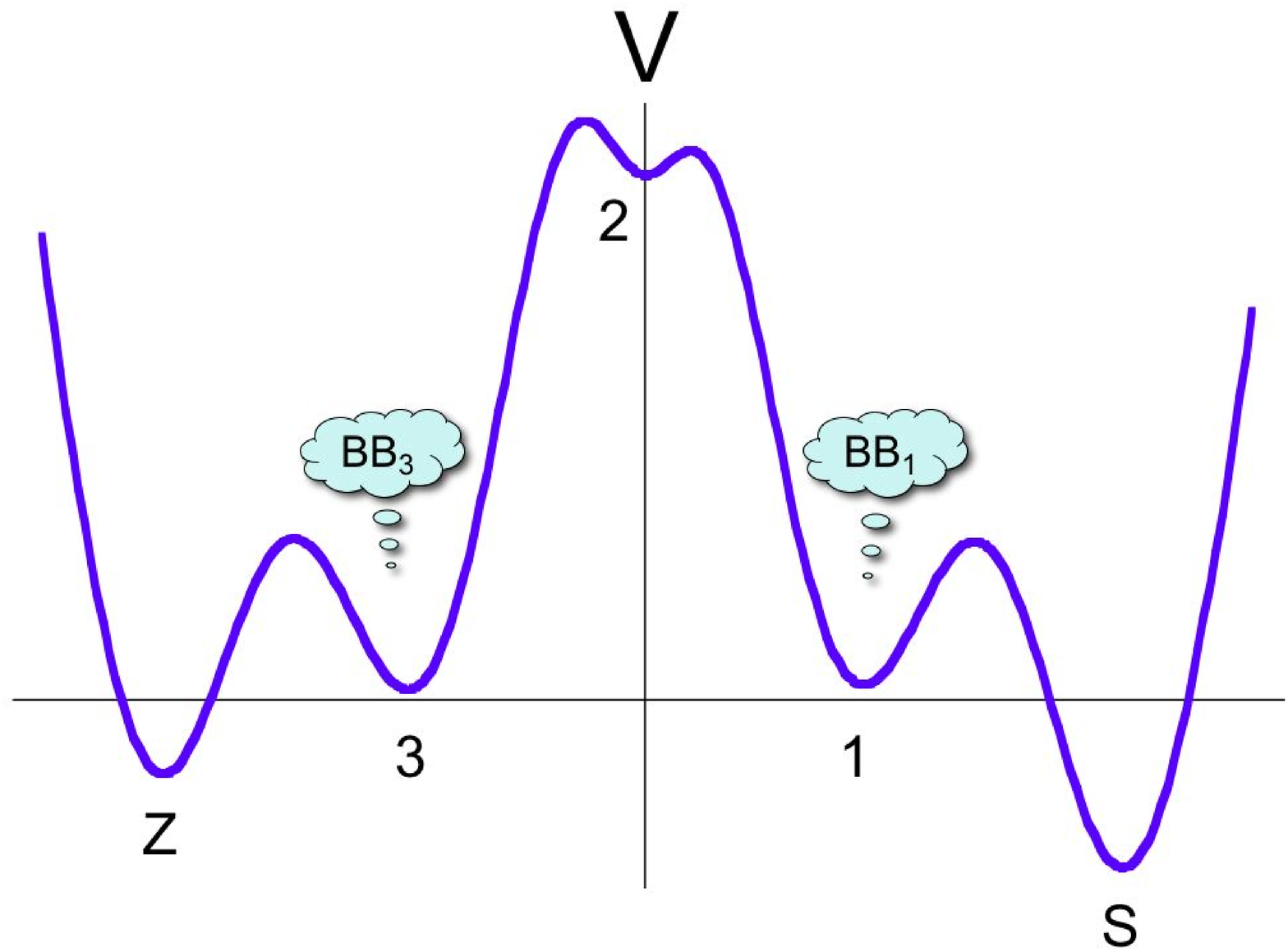} \caption{Boltzman brains populating  the landscape}\label{BBfig}
\end{figure}

Let us consider now a more complicated regime, with the potential having three different dS states and two AdS sinks, Fig. \ref{BBfig}. Suppose again that we began in the upper dS. But now we can either fall to the right or to the left. If you fall to the right, to the wide sink, you live there for a short time and die without being reborn as a BB. But if the probability to fall to the left is bigger, and if the left sink is narrow (low probability to decay to the sink), then you are going to be a BB. To avoid this problem, decay probabilities of all low dS vacua must be very high \cite{Bousso:2006xc}. Whilst this is possible, it is a strong constraint on the string theory landscape, which may or may not be satisfied.

But the most interesting feature is the same as in the one-sink model: The lower we begin, the higher is the probability to become a BB.  This means in particular, that if  the Hartle-Hawking wave function \cite{Hartle:1983ai} describes creation of the universe from nothing, then the probability that the universe is created in the upper dS vacuum is exponentially suppressed. A typical universe should be born in the lowest dS vacuum, and therefore it becomes populated  by BBs, even if all sinks are wide. But  if creation of the universe is described by the tunneling wave function \cite{Linde:1983mx}, then the universe is created in the highest dS space, and the chances that it will be populated by normal people will be much higher, though still not guaranteed.

On the other hand, one may argue that quantum creation of a {\it compact}  flat or open inflationary universe (e.g. a toroidal universe) is much more probable than the quantum creation of a closed universe studied in  \cite{Hartle:1983ai,Linde:1983mx}. This process may occur without any exponential suppression, practically independently of the initial value of the effective potential  \cite{Linde:2004nz} (see also \cite{Zeldovich:1984vk}). Then one can start inflation at any maximum or minimum of the effective potential with almost equal ease. This means that an additional investigation is required to integrate over all possible initial conditions discussed above and find the actual predictions of the scenario based on the local description.

\subsection{BBs and the pseudo-comoving volume-weighted distribution}\label{cer}

One can perform a similar analysis in the pseudo-comoving volume-weighted distribution discussed in Sect. \ref{CVW}, see \cite{Ceresole:2006iq}. In this case the final results do not depend on initial conditions, for the reasons discussed in Sect. \ref{CVW}.

The final result for the two dS vacuum case is
\begin{equation}
\label{qbbp} {Q_{BB}\over Q_{1}} =  {P_{1}\Gamma_{1B}\over P_{2} \Gamma_{21}}  = {\Gamma_{1B}\over \Gamma_{1s}} \   ,
\end{equation}
under the condition $\Gamma_{1s} \gg \Gamma_{21}$. Thus one does not have the BB problem for $ \Gamma_{1s} \gg  \Gamma_{21}, \Gamma_{1B}$.

In general,   the condition $ \Gamma_{1s} \gg   \Gamma_{1B}$ can be quite restrictive.  If we consider many vacua and the required conditions are not satisfied near some of them, then the corresponding BBs may dominate.

\subsection{BBs and the  standard volume-weighted distribution}\label{st}

So far we studied the local description \cite{Bousso:2006xc}, and the global description  \cite{Ceresole:2006iq}, and both of them have demonstrated rather mixed results in solving this problem.

The reason is very easy to understand: In both cases we took some pain not to reward exponentially growing parts of the universe for producing lots of space very quickly. Now let us turn our attention to the distribution discussed in Sect. \ref{VW}, which takes into account different rates of growth of volume of the different parts of inflationary domains. In this case, using the results obtained in Sect. \ref{VW} one easily finds that
\begin{equation}
\label{qvwbbp} {Q_{BB}\over Q_{1}} =      {P_{1}\Gamma_{1B}\over P_{2} \Gamma_{21}} =  {\Gamma_{1B}\over 3(H_{2}-H_{1})}   \   .
\end{equation}
If one takes the typical estimate of the rate of the BB production $\Gamma_{1B}\sim 10^{-10^{50}}$ \cite{Page:2006dt,Bousso:2006xc}, and compares it with any reasonable value of $H$, from the Planck value $O(1)$ to the present value $10^{-60}$, one can easily see that the relative probability to be a BB in this approach is given by
\begin{equation}
\label{qvwbbp1} {Q_{BB}\over Q_{1}} \approx \Gamma_{1B}\sim 10^{-10^{50}}  .
\end{equation}
This completely solves the problem, and this solution does not depend on initial conditions,  on the wave function describing quantum creation of the universe, or on the existence of the sinks in the landscape.

The same solution will remain valid for any potential $V$, however complicated it may be. Indeed, the main feature of this probability distribution \cite{LLM} is that the growth of the physical volume of our universe mostly occurs due to the growth of  domains with the largest values of the Hubble constant. Then the parts of the growing volume in the highest dS vacuum tunnel  down, and produce observers like ourselves. And only a tiny part of this flux, proportional to $\Gamma_{1B}\sim 10^{-10^{50}}$,  turns back due to quantum fluctuations, and produces Boltzmann brains. That is why we have never seen them.

 Before concluding this section, let us discuss a generalized volume-weighted probability measure, which leads to the following equations:
\begin{equation}
\label{v1vw00} \dot P_{1} =  -P_{1}\, (\Gamma_{1s} +\Gamma_{12}) +  P_{2}\,\Gamma_{21} +3 H_{1}^{\beta}P_{1}\ ,
\end{equation}
\begin{equation}
\label{v2vw00} \dot P_{2} =  -P_{2}\, (\Gamma_{2s} +\Gamma_{21}) +  P_{1}\,\Gamma_{12} +3  H_{2}^{\beta}P_{2}\ .
\end{equation}
Parameter $\beta$ corresponds to different choices of time parametrization: The time can be measured in units $H^{\beta-1}$. The case $\beta = 0$ describes the pseudo-comoving probability distribution, $\beta = 1$ corresponds to the `standard' probability distribution, see e.g. \cite{Garriga:2005av}.  

One can easily check that this probability measure does not suffer from the BB problem unless one considers the limiting case $\beta \ll \Gamma_{1B} \sim 10^{-10^{50}}$. In other words, the BB problem does not appear in a broad class of the volume-weighted distributions, unless one considers a special limiting case $\beta \to 0$ corresponding to the pseudo-comoving measure described in the previous section.

\section{The standard volume-weighted distribution and the cosmological constant problem}\label{CC}

One of the main reasons of the recent interest in the string theory landscape is the possibility of the anthropic solution of the cosmological constant (CC) problem. Therefore it is important to check whether the proposed solutions of the BB problem are compatible with the anthropic solution of the CC problem.  

The first anthropic solution of the cosmological constant problem in the context of inflationary cosmology was proposed back in 1984 \cite{Linde:1984ir}. It was based on the local description of the universe and on the  calculation of the probability of initial conditions using the tunneling wave function \cite{Linde:1983mx}. It was assumed there that the vacuum energy density is a sum of the scalar field potential $V(\phi)$ and the energy of fluxes $V(F)$. According to  \cite{Linde:1983mx}, quantum creation of the universe is not suppressed if the universe is created at the Planck energy density, $V(\phi) +V(F) = O(1)$, in Planck units. Eventually the field $\phi$ rolls to its minimum at some value $\phi_0$, and the vacuum energy becomes $\Lambda = V(\phi_0) + V(F)$. Since initially $V(\phi)$ and $V(F)$  could take any values with nearly equal probability, under the condition $V(\phi) +V(F) = O(1)$, we get a flat probability distribution to find a universe with a given value of the cosmological constant after inflation, $\Lambda = V(\phi_0) + V(F)$, for $\Lambda \ll 1$. The flatness of this probability distribution is crucial, because it allows us to study the probability of emergence of life for different $\Lambda$. Finally, it was  argued in   \cite{Linde:1984ir}  that life as we know it is possible only for  $|\Lambda|  \lesssim \rho_{0}$, where $\rho_{0} \sim 10^{{-120}}$ is the present energy density of the universe.  This fact, in combination with inflation, which makes such universes exponentially large, provided a possible solution of the cosmological constant problem. 

The anthropic constraint $ -\rho_{0} \lesssim \Lambda  \lesssim   \rho_{0}$ was used in several other papers on this subject \cite{Davies,Sakharov,Banks,300}. The first part of this constraint, $-\rho_{0}  \lesssim \Lambda$, was quite trivial. Indeed, for large negative $\Lambda$ the universe would  collapse before life could emerge there    \cite{300,Barrow,Kallosh:2002gg}. Meanwhile, the justification of the upper bound was much more complicated; it was derived in the famous papers by Weinberg  \cite{Weinberg:1987dv} and subsequently improved in a series of papers by other authors, see e.g. \cite{Garriga:1999hu,Tegmark:2005dy}.  The final result of these investigations, $|\Lambda|  \lesssim O(10)\rho_{0} \sim 10^{-119}$, is very similar to the bound used in \cite{Linde:1984ir}.

The simple model proposed in  \cite{Linde:1984ir},  was based on the assumption that the choice between various values of $V(F)$ occurs only at the moment of the quantum creation of the universe, because at the classical level the field $F$ must be constant.  The situation becomes more complicated if one considers quantum  transitions between different dS flux vacua  \cite{Bousso:2000xa}, because eventually (within the time $t \sim  e^{24\pi^{2}/\Lambda}$) the probability distribution $P$ becomes proportional to the Hartle-Hawking distribution $e^{24\pi^{2}/\Lambda}$ (in the absence of wide sinks), see Section \ref{dS}. Naively, one could expect that this means that the cosmological constant must take its smallest possible value. However, as we already emphasized, in anthropic considerations one should use the probability currents instead of the probability distribution $P$ \cite{Bellido}. The stationary probability distribution $e^{24\pi^{2}/\Lambda}$ is established when the currents upwards are equal to the currents downwards, which implies  that the probability currents between two different dS states do not depend on the value of the cosmological constant in each of these states.

This fact  by itself is insufficient for the flatness of the prior probability distribution for the cosmological constant in the landscape. Indeed, if we have more than two vacua, then the incoming currents to different dS vacua in general will be different, despite the overall balance of currents. Therefore one may wonder whether the probability distribution for different values of $\Lambda$ is flat, as  required  for the derivation of the anthropic constraints on $\Lambda$. The situation becomes even more complicated in the presence of wide sinks, or if one studies volume-weighted probability distributions. 

However, one may argue that if the total number of dS vacua is extremely large, then, on average, the probability flux to the set of all vacua with $\Lambda \ll 1$ should not depend on $\Lambda$.\footnote{I am very grateful to Lenny Susskind who emphasized this point to me.} In the context of the string theory landscape, this situation was analyzed in \cite{Perlov}, using the probability measure of Ref. \cite{Vilenkin:2006qf}. In the model with $10^{7}$ vacua, they found a staggered (i.e. very non-flat) probability distribution for $\Lambda$, but argued that it may become sufficiently smooth and flat if the number of different vacua $\cal N$ is sufficiently large. Here we will try to develop this argument further and see what may happen if we use the `standard' volume-weighted probability distribution, which just saved us from the BB problem. 

As we have seen in Sect. \ref{VW}, the main feature of the `standard' probability measure is that the probability distribution (the fraction of the total volume) $P_{i}$ in all vacua grows at the same rate. This rate is determined by the Hubble constant $H_{h}$ in the highland, which is  the set of the highest dS vacua in the string theory landscape,  $P_{i} \sim e^{H_{h} t}$. This result is very similar to the result obtained in \cite{LLM} for  slow-roll eternal inflation. Equal rates of growth of all parts of the universe occurs due to a powerful probability current, which flows  from the highland to all other dS minima. We argued in Section  \ref{VW} that the ratio of the probability currents from the highland to the lower minima $dS_{i}$ is proportional to 
 $\Gamma_{hi}$, where  $\Gamma_{hi}$ describes the probability of the transition from the highland to the $dS_{i}$ vacuum,   see Eq. (\ref{highlanddown}):
 \be\label{ccdistr}
 P(\Lambda_{i}) \sim  \Gamma_{hi} \ .
 \ee
 
 If the total number of dS vacua is not very large, this may lead to a staggered probability distribution similar to the distribution found in \cite{Perlov}. On the other hand, in the realistic situation with $10^{500}$ (or perhaps even $10^{{1000}}$) different vacua, the probability distribution can be quite smooth, and it may be flat for $\Lambda$ in the anthropic range of its values. While we cannot present a rigorous proof of this conjecture, we will present here three different arguments which point in this direction.

1) The value of $\Lambda$ after the uplifting is a sum of the depth of one of the $10^{500}$ AdS minima before the uplifting, $V_{\rm AdS}<0$, and  of the uplifting term,  which usually cannot be much greater than  $|V_{\rm AdS}|$, because otherwise it may destabilize the volume modulus \cite{Kallosh:2004yh}.   Thus,   after the uplifting $\Lambda$ can be either positive or negative; typically it should be somewhere in the interval  $|\Lambda| \lesssim V_{\rm AdS}$. Since the typical value of $|V_{\rm AdS}|$ can be a hundred orders of magnitude higher than the anthropic range of values of the cosmological constant, one may argue that the distribution of values of  the cosmological constant in the anthropically allowed range $|\Lambda| \lesssim 10^{-119}$ should not depend on the precise value of $\Lambda$; see e.g. \cite{Douglas:2006es}.

2) Among all possible tunneling trajectories, the only ones that are anthropically allowed are the trajectories that end up with a long stage of  slow-roll inflation, reheating and baryogenesis. If these conditions are not satisfied, the interior of the corresponding bubble 10 billion years after its formation will be an empty open universe. To produce baryons, one may need to have reheating temperature greater than 1 GeV, which requires the vacuum energy $V$ prior to the last tunneling event to be greater than $10^{-70}$, in Planck units. This may not seem to be a very big number, but it is 50 orders of magnitude greater than $10^{-120}$.

Note that the relatively large amplitude of the last jump is in fact quite natural. Indeed, let us assume, following  \cite{Perlov}, that each time  the tunneling occurs, we can make only one step in the lattice, by jumping to the neighboring flux vacuum. In such a situation one may expect that the vacuum energy changes by some fraction of $|V_{\rm AdS}|$, which is dozens of orders of magnitude greater than $10^{-120}$. Therefore  any jump to the  vacua in the anthropically allowed range $|\Lambda| \lesssim 10^{-119}$ should occur from much higher vacua. In this case, the precise value of $\Lambda$ in the anthropic  range cannot influence  the tunneling rate $\Gamma_{hi}$. In other words,  the tunneling rates $\Gamma_{hi}$ along all anthropically allowed tunneling trajectories can be very sensitive to the change of various parameters of the theory, but these rates, on average,  cannot have an explicit dependence on  $\Lambda$ in the anthropic  range of its values.

Since neither the distribution of $\Lambda$ in different dS vacua nor the probability of tunneling to these vacua are expected to depend on  $\Lambda$ in the anthropically allowed range of its variations, we should have a flat probability distribution for $\Lambda$. The remaining issue is to check whether this distribution is smooth. More exactly, one should check how large should be the total number of the vacua to achieve smoothness; we may have $10^{500}$ or $10^{1000}$ different vacua, but perhaps not $10^{10^{120}}$.\footnote{The maximal number of the vacua in the landscape is  not known  yet \cite{Douglas:2006es}. Also, one can have an effectively continuous distribution of the vacua if we have a potential with an extremely small slope, as in \cite{Banks,300}. I am not sure whether one can find a realization of this situation in string theory.}

3) The smoothness of the distribution  is the most complicated part of the problem; let us try to analyze it by making some model-dependent estimates. Consider, for example, a lattice of the flux vacua, of the type described in \cite{Bousso:2000xa}. Imagine that it is a N dimensional lattice box, of a size L, so that the total number of vacua is ${\cal N} \sim L^{N}$. Suppose that we need to tunnel from one corner of this box (highland) to another (our vacuum with $\Lambda \sim 10^{{-120}}$). We will assume, as before, that each time  the tunneling occurs, we can make only one step in the lattice, by jumping to the neighboring flux vacuum   \cite{Perlov}. We will also assume, following \cite{Kachru:2003aw}, that the rate of such tunneling from the state with  entropy ${\bf S_{i}} = {24\pi^{2}\over V_{i}}$ cannot be suppressed by more than $e^{-\bf S_{i}}$, and we will follow each tunneling trajectory until it brings us close to the anthropic range of $|\Lambda| = O(10^{-119})$. We will assume that the tunneling mainly goes down, because the probability of the jumps upwards is strongly suppressed. To make a numerical estimate, we will assume  also, in accordance with the previous arguments, that the main part of the process occurs at $V > 10^{-70}$.  Then one may expect that the maximal combined suppression of the probability to tunnel down along this trajectory cannot be stronger than $\Gamma \sim \bigl[\exp (-10^{70})\bigr]^{NL} = \exp (-NL\times {10^{70}})$. For example, for $L = 100$, $N = 250$, one has ${\cal N} \sim 10^{500}$ and $\Gamma_{m}\sim  \exp (-2500\times {10^{70}}) \sim \exp (- {10^{74}})$. 

This should be compared to another extreme limit: If  all jumps are  suppressed only marginally, e.g., by a factor of $e^{-O(1)}$, and only one jump is required to bring us down, one can get $\Gamma  = e^{-O(1)}$.
 
Naively, one could worry that in order to find a smooth distribution of the possible values of the cosmological constant (\ref{ccdistr}), one would need much more than $10^{500}$ vacua to compensate for the incredibly broad dispersion of the decay rates, from $\Gamma \sim \exp (- {10^{74}})$ to $\Gamma \sim e^{-O(1)}$. However,  the rate of tunneling $\Gamma_{hi}$ depends  exponentially on the physical parameters of the theory, such as coupling constants, or the dS entropy, which is inversely proportional to the vacuum energy. If one assumes that the distribution of these parameters is relatively uniform, then one  can expect a certain uniformity of the distribution of $\log^{{-1}}   \Gamma_{hi}$. In this case one can divide the total interval of all possible values of  $\log^{{-1}}   \Gamma_{hi}$ to the bins of size 1, as   in \cite{Perlov}, and argue that $10^{500}$ different values of $\log^{{-1}}   \Gamma_{hi}$ are relatively evenly distributed among these bins, so that each bin will correspond to approximately $10^{500-74} = 10^{426}$ different decay channels.  On the other hand, if the coupling constants and the values of dS entropy  depend exponentially on the values of the fluxes etc., and these values in  turn are uniformly distributed, then the total number of bins will be even  smaller, and the total number of vacua in each bin will be even greater.

One can expect that among $10^{426}$ different decay channels in each bin, $10^{426-120} = 10^{306}$ channels will bring us to the vacua close to the anthropic range of $|\Lambda| = O(10^{-120})$. Since all of these trajectories will belong to the same bin, the tunneling rates along all of them will be comparable. This suggests that the distribution of the  rates of decay to the  vacua close to the anthropic range of $|\Lambda|$ can be nearly continuous, and  there will be exponentially many  anthropically allowed vacua which belong to the same bins and therefore  cannot be distinguished from each other by the corresponding values of  $\Gamma_{hi}$. This implies that the prior probability of different values of $\Lambda$ close to their anthropically allowed range is expected to be continuous and flat.

Our argument was based on many assumptions; it is plausible but not water-tight. 
To begin with, all examples of the tunneling in the landscape studied so far were based on the investigation of the situations where either scalar fields change, or fluxes change, but we need to study the situation where these changes occur simultaneously, because the values of the scalar fields depend on fluxes. The estimates given above may change if we concentrate on the subclass of the  vacua that can describe inflation and the standard model. This may increase the total number of the vacua required for the smoothness of the distribution, but we do not see any reason to expect that the required numbers should be as large as $10^{10^{120}}$; our  estimates indicate that $10^{500}$ or $10^{1000}$ vacua may be quite sufficient.

Our main arguments were based on the flatness of the distribution of the vacua with different $\Lambda$, on the $\Lambda$-independence of  the probability of tunneling  from dS vacua of a much greater height, and on the large number of vacua in the string theory landscape. All of these features of the landscape are quite independent  of the choice of the measure in the multiverse.  
It is quite possible that the required number of the vacua, which ensures the smoothness of the distribution, will depend on the choice of the measure.  However, our investigation suggests that if the total number of stringy vacua is sufficiently large,  the anthropic solution of the cosmological constant problem  may be valid independently of the choice of the measure. This may reduce the sensitivity of our predictions to the as yet unsolved problem of the choice of the measure, at least with respect to the cosmological constant problem. This is a very important issue which deserves a separate detailed investigation.


\section{Discussion}\label{disc}

In this paper we considered three simplest probability measures
discussed in the literature. The first one is a comoving probability
distribution, which follows the evolution of individual points,
ignoring the fact of expansion of the universe, see Section
\ref{comoving}. The second one is very similar, in this paper we
called it quasi-comoving, see Section \ref{CVW}. It does not reward
any parts of the universe for the different rate of their expansion,
but it calculates ratios of different fluxes and volumes, keeping in
mind that the total number of `points' during eternal inflation
grows exponentially. One can think about it as the probability
measure which appears when one studies the physical volume of
different parts of the universe at the hypersurface of equal time,
when the time is measured in units of $H^{{-1}}$ along each
geodesic. The third probability measure appears when one studies the
physical volume of different parts of the universe at the
hypersurface of equal time,  measured in units of $M_{p}^{{-1}}$, or the string time $M_{s}^{-1}$
see Section \ref{VW} For lack of a better word we called this
probability measure `standard,' which emphasizes the fact that it uses the standard physical time measured in the number of oscillations rather than  the very unusual time measured in units of $H^{{-1}}$. The standard probability measure takes into account the
difference between the rate of expansion in different parts of
inflationary universe. For each of these three cases we defined the
incoming probability current, which is most suitable for anthropic
applications.

There are several other probability distributions discussed in the
literature. One of the most sophisticated is the proposal described
in \cite{Vilenkin:2006qf}. The reason why we concentrated on the
three `toy model' measures described above is that they are
relatively simple, and all of them can be formulated in a unified
way, so they can be easily analyzed and compared to each other.
Moreover, some of these measures enter as a part of other, more
complicated proposals discussed in the literature. When we study
these three proposals, we can learn a lot, do it  quickly, and then
we can use our experience in discussing other options.

Investigation of the comoving probability distribution shows that
its properties depend in a very important way on the existence of
terminal vacua, which we called sinks in the landscape. In the
presence of the sinks, which is a generic property of the string
theory landscape, this approach by itself does not allow us to make
definite predictions, so it should be  supplemented by the theory of
initial conditions for inflation.

Predictions of the probability measure based on the quasi-comoving
probability distribution do not depend on initial conditions, since
they are quickly forgotten in the course of eternal inflation.
However, these results are very sensitive to the existence of the
sinks. We identified two different regimes, which we called `narrow
sink' and `wide sink' regimes. In the narrow sink case, the
probability distribution remains effectively thermalized, as in the
eternal de Sitter case, whereas in the presence of wide sinks one
encounters non-thermalized probability currents
\cite{Ceresole:2006iq}.

Finally, for the `standard'  probability distribution, rewarding
fast growers, nothing depends on initial conditions and on the
sinks, because of the powerful probability current which flows from
the dS regions with the highest possible values of the Hubble
constant \cite{LLM,Bellido}.

As a test of these probability distributions, we studied the
Boltzmann brain  problem recently discussed in the literature. One of
our goals was to verify the conjecture that all probability measures
based on the global description of the universe  suffer from the BB
problem  \cite{Bousso:2006xc}.   As we have seen, the comoving
probability distribution, based on the local description of the
universe, as well as the pseudo-comoving probability distribution,
based on the global description but not rewarding fast growers, are
potentially vulnerable   to the  BB problem. Meanwhile, the
`standard' volume-weighted probability measure, which is one of the
simplest and most natural volume-weighted probability distributions
proposed in \cite{Eternal,LLM,Bellido}, completely solves the BB
problem.

One may consider this fact as an argument in favor of the `standard'
probability measure proposed in \cite{Eternal,LLM,Bellido}. On the
other hand, it is quite possible that other probability measures can
be equally successful. For example, it is quite possible that one can solve the BB problem in the context of the approach of Ref. \cite{Vilenkin:2006qf}. 

To analyze this case, one should note that the Boltzmann brains cannot really live in the vacuum. There should be some structures which make life possible. These structures include bubbles, galaxies, or planets. It is natural to divide the process into two parts: producing structures and producing life. BBs are the observers living in the galaxies produced due to thermal fluctuations, normal observers live in the galaxies produced in a more regular way, after inflation inside the bubbles.
Producing galaxies or planets due to thermal fluctuations in dS space in some cases  may be more probable than producing bubbles.  But the difference in probabilities is finite. Meanwhile each bubble contains infinitely many observers, so one may argue that normal people should beat BBs hands down. This argument is very similar to the argument recently proposed by Vilenkin \cite{Vilenkin:2006qg}.

It is quite important that not all probability measures pass the BB test. For example, the  probability measure recently proposed by Don Page \cite{Page:2006ys} fails this test.  Therefore one may argue that this measure   is ruled out by observations. This is not surprising, since the measure proposed in  \cite{Page:2006ys} ignores all observers which are born after the new bubble production. I find this condition very hard to justify.

In what follows we will discuss some problems associated with the
`standard' probability measure, their possible resolution, and the
situation with anthropic predictions in general.

1) The volume-weighted probability distributions discussed in our
paper give predictions that depend on the choice of the time
coordinates used to perform the time slicing (measuring time  in
units of $H^{-1}$, or in units of $M_{p}^{{-1}}$ or  $M_{s}^{-1}$)
\cite{LLM,Bellido}. However, as we already argued, all local processes, including chemical and biological processes, are most properly described in standard time, measured in units of $M_{p}^{{-1}}$,  or  $M_{s}^{-1}$, instead of the time measured in units of  $H^{-1}$, which describes the logarithm of the distance between different galaxies.  One may use this fact as an argument in favor of  the standard probability measure.

2) This measure predicts unusual nonperturbative corrections to the
large scale structure of the universe. If these corrections are
large, then we may find ourselves near the center of a spherically
symmetric bubble \cite{Linde:1996hg}. Some theorists consider it
undesirable, whereas some observers argue that we may actually live
in one of these bubbles. One way or another, this effect is small
and unobservable if inflation  does not occur at a nearly Planckian
density  \cite{Linde:1996hg}. In particular, no effects of such type
are possible in new inflation, hybrid inflation, and all known
versions of inflation in string theory.

3) If one uses the standard probability distribution $P$ for predicting
the part of the universe where we should live, one may conclude that
the universe surrounding us must be very young (the youngness paradox \cite{Guth:2000hz}) and unacceptably hot
\cite{Tegmark:2004qd}. However, as we already emphasized, instead of
using the probability distribution $P$, in anthropic
considerations  one should use the incoming probability currents $\dot Q$, or the integrated probability currents $Q$ 
\cite{Bellido}. If one does so, the paradoxes discussed in   \cite{Guth:2000hz,Tegmark:2004qd} disappear.

We can continue discussing possible problems of various
volume-weighted probability measures
\cite{Mediocr,Feldstein:2005bm,Garriga:2005ee},  and their
resolution
\cite{Garcia-Bellido:1994ci,Linde:2005yw,Lyth:2005qk,Hall:2006ff},
but we will stop here before we forget that the `standard' volume
weighted measure  \cite{Eternal,LLM,Bellido} just saved us from the
BB disaster.

In addition to testing various probability measures, one may also
try to understand whether they are competing with each other, or
complementing each other, because each of  them can be useful for
answering different types of questions.

Let us give a simple example related to demographics. One may want
to know what is the average age of a person living now on the Earth.
In order to find it, one should take the sum of the ages of all
people and divide it by their total number. Naively, one could
expect that the result of the calculation should be equal to $1/2$
of the life expectancy. However, the actual result will be much
smaller. Because of the exponential growth of the population, the
main contribution to the average age will be given by very young
people. Both answers (the average age of a person, and a half of the
life expectancy) are correct despite the fact that they are
different. None of these answers is any better; they are different
because they  address different questions.  Economists may want to
know the average age in order to make their projections. Meanwhile
each of us, as well as the people from  the insurance industry, may
be more interested in the life expectancy.

Similarly, all possible ways to measure our universe may be useful
for answering different questions. The comoving probability
distribution can be used to study a typical evolution of the
physical conditions at a given point. Meanwhile,  the global
structure of the universe  can be studied using various
volume-weighted distributions. The problem appears only if one wants
to find out which of these probability distributions, if any, should
be used in   anthropic considerations.

There are several ways to approach this problem. One of them, which
we already discussed and used in this paper, is to compare
predictions of each of these measures with observations. In this
sense, the probability measure becomes a part of the theory, and we
test both the theory and the measure by comparing them with
observations.

Another strategy (which may be used in parallel with the first one)
is to reduce metaphysical overtones of the anthropic principle by
asking well defined questions about conditional probabilities and
treating all available facts, including the facts related to our
life,   as observational data.  If one does not do it, one may come
to all sorts of paradoxical conclusions.

For example, one may wonder what is the most probable state of the
universe compatible with life if one allows all possible parameters
to vary, e.g. if one can vary the cosmological constant and   the
amplitude of density perturbations  simultaneously
\cite{Garriga:2005ee}. The results of this approach can be  rather
problematic, but several solutions to this problem are available
\cite{Garriga:2005ee,Linde:2005yw,Lyth:2005qk}.  Considerable
ambiguities appear in imposing  the anthropic constraints on the
amplitude of spontaneous symmetry breaking in the standard model if
one can vary all possible parameters of the standard model and
beyond \cite{Harnik:2006vj}; see, however,
\cite{Feldstein:2006ce,Clavelli:2006di}. Yet another example is the
possibility to predict just about any value of the cosmological
constant if one considers information processing in the universe and
does not specify what kinds of observers are making the
observations, i.e. if one ignores the word `anthropic' in the
`anthropic principle' \cite{Starkman:2006at}.

One may argue that the main root of some of these problems is that
we want to achieve too much. Before the invention of inflationary
cosmology and string theory landscape, we wanted to construct a
unique theory explaining all features of our world.  Now we
understand that the world may consist of many different parts and we
can live only in some of them, but we want  to replace the idea that
our universe is unique by a closely related hypothesis that our
position is most probable, that we live in a state where a typical
observer should live, etc. This is a very powerful idea, and one
should pursue it as far as possible. But it resembles a question
which has bothered me for a long time: Why was I born in Moscow if
there are so many other places, and many more people live in those
places? And why was I born in the middle of the 20th century, if
the population grows exponentially, and  many more people are alive
now than they were at the time when I was born?

Unfortunately, there is a chance that at least some of these
questions are meaningless, or we are not ready to answer them, and
we will be able to do it only after we learn much more, not only
about the nature of the universe, but also about the nature of life.
Thus, these questions may remain unanswered for a long time, but it
does not preclude us from answering simpler, more pragmatic
questions, based on the conditional probabilities. For example, even
if one cannot calculate the probability to be born in some
particular place at some particular time, one may ask questions
about the most probable observational results under the condition
that one is born there. If I am born in Moscow, and everybody around
speaks Russian, it is not surprising. But  if I find that  everybody
in Moscow speaks Chinese, I will be really surprised, and I will try
to come with some theoretical explanation.

Let us try to use this analogy and apply it to our use of anthropic
considerations. The first observations which we make give us some
primitive information about our environment, about other people,
about the city and the country where we were born. Then we learn
that our bodies are mostly made from hydrogen, oxygen and carbon. At
this stage we do not yet know  about the cosmological constant and
the Higgs field. But we already know that among all possible parts
of the universe {\it we}\,  (rather than some generic observers) can
live only in the parts which can support {\it carbon-based life}
(rather than  silicon-based life, or some life-like information
processing in general). When evaluating the conditional
probabilities of the results of future observations, we must take
into account that  {\it we} are going to make these observations.
This implies that we are talking about {\it anthropic} principle,
rather than about  {\it atomic} principle, or  {\it galactic}
principle.  We are doing what all of us consider quite legitimate:
We evaluate the probability of the outcome of  future observations
on the basis of the previously obtained data. The only nontrivial
step here is that we consider simple facts of our life (like the
fact that we are alive and made from hydrogen, oxygen and carbon) as
observational data. This may seem unusual at first, but it is no
more unusual than considering the facts that our universe is big and
parallel lines do not intersect as observational data. In the end,
this strategy lead us to the discovery of inflation.

Let us apply it to the cosmological constant problem. A long time
ago, we already knew that the amplitude of density perturbations
required for the formation of galaxies was about $10^{{-5}}$. Later
we learned that the cold dark matter scenario makes a better job in
describing  formation of galaxies than the hot dark matter scenario.
We did not know yet what was the vacuum energy, and the prevailing
idea was that we did not have much choice anyway. But with the
discovery of inflation, we learned that the universe could be
created differently, with different values of the cosmological
constant in each of its copies described by the different branches
of the wave function of the universe,  or in each of its parts
created by eternal inflation. This allowed us to propose several
different anthropic solutions to the cosmological constant problem
based on the assumption that,
for the given value of the amplitude of density perturbations, we
cannot live in a universe with $|\Lambda| \gg 10^{-120}$. 
If observations would show that the
cosmological constant vanishes, or if it were a thousand times
smaller than the anthropic bounds, then we would be surprised, and a
theoretical explanation of this anomaly would be required. As of
now, the small value of the cosmological constant does not look too
surprising, but we are still working to understand the situation
better.

Historically, in our analysis we did not try to explain all constants at
once by finding the best place in the universe populated by most of
the observers capable of information processing. Instead of that, we
operated using the standard rules of science. We found the
conditions necessary for the existence of life of {\it our } type,
we measured   the value of the amplitude of density perturbations and the gravitational constant in
{\it our } part of the universe, and only after that we  evaluated
the possible range of the cosmological constant consistent with all
previously known experimental data describing the part of the
universe where {\it we} live.

The situation with the anthropic bounds on the amplitude of the
Higgs field   \cite{Agrawal:1997gf,Arkani-Hamed:2004fb,Hogan:2006xa} is very similar. Long ago, we
measured masses and coupling constants of many elementary particles,
we found some basic facts about the weak interactions, but for a
while we did not know the amplitude of the Higgs field responsible
for the electroweak spontaneous symmetry breaking. Then we found
anthropic constraints on the value of the Higgs field {\it in our
part of the universe}, under the condition that all other parameters
are fixed, because we already measured them  {\it in our part of the
universe}. This procedure does not involve any of the ambiguities
described in \cite{Harnik:2006vj}.

This does not mean that we should not go beyond this simple
step-by-step procedure.  We should certainly continue our attempts
to understand everything from first principles, and maybe eventually
we will find that our presence in this part of the universe with all
of its bizarre  properties was most probable. This would be a great
triumph of science. That is why we should continue working, trying
to find an ultimate explanation of the physical reality, despite all
problems and uncertainties which we encounter. But on the other
hand, perhaps one should not be too disappointed if eventually we
find that at least some of our pre-Copernican ideas are  wrong, and
we do not occupy the central, or typical, or most probable position
in the world. Even in this case the anthropic principle will find a
decent place in the theorists toolbox, by allowing us to pre-select
those parts of the universe  and those vacua in the landscape where
{\it we} can live.

And there is still another possibility. It may happen that some of the predictions of our calculations will not be sensitive to the choice of the probability measure. (A similar hope was expressed long ago in \cite{Garriga:1997ef}.)  For example, the arguments contained in Sect. \ref{CC} indicate that the cosmological constant problem can be solved using the `standard' probability measure \cite{LLM,Bellido}. The main ingredients of the proposed solution are measure-independent; they depend only on the properties of the landscape.  This is quite encouraging, since it suggests that under certain conditions the solution of the cosmological constant problem may be measure-independent. It would be very important to verify this conjecture, because it would remove a lot of uncertainties related to the choice of the measure in the string landscape scenario.

\

\noindent{\large{\bf Acknowledgments}}

\noindent
It is a pleasure to thank A.~Aguirre,  R. Bousso, T. Clifton,  B. Freivogel, J.~Garriga, M. Graesser, S.~Gratton, M. Johnson, R. Kallosh,  D. Page, M. Salem, S. Shenker,  L. Susskind, N. Sivanandam,  A. Vilenkin and S.~Winitzki for many stimulating discussions.  This work   was supported by NSF grant PHY-0244728.

\

\end{document}